\documentclass[10pt,aps,twocolumn,prd,notitlepage,amssymb,amsmath,floatfix,nofootinbib,superscriptaddress]{revtex4-1}
\usepackage{epsfig}
\usepackage{bm}
\usepackage{amssymb}
\usepackage{amsmath}
\usepackage{color}
\usepackage{xcolor}
\usepackage{slashed}
\usepackage[colorlinks,linkcolor=blue,anchorcolor=black,citecolor=blue]{hyperref}
\usepackage{multirow}
\usepackage{ulem}
\usepackage{comment}

\allowdisplaybreaks[4]

\begin{document}

\title{Longitudinal and transverse polarizations of $\Lambda$ hyperon in unpolarized SIDIS and $e^+e^-$ annihilation}

\author{Kai-bao Chen}
\email{chenkaibao19@sdjzu.edu.cn}
\affiliation{School of Science, Shandong Jianzhu University, Jinan, Shandong 250101, China}

\author{Zuo-tang Liang}
\email{liang@sdu.edu.cn}
\affiliation{Institute of Frontier and Interdisciplinary Science, 
Key Laboratory of Particle Physics and Particle Irradiation (MOE), Shandong University, Qingdao, Shandong 266237, China}

\author{Yu-kun Song}
\email{sps\_songyk@ujn.edu.cn}
\affiliation{School of Physics and Technology, University of Jinan, Jinan, Shandong 250022, China}

\author{Shu-yi Wei}
\email{shuyi@sdu.edu.cn}
\affiliation{Institute of Frontier and Interdisciplinary Science, 
Key Laboratory of Particle Physics and Particle Irradiation (MOE), Shandong University, Qingdao, Shandong 266237, China}
\affiliation{European Centre for Theoretical Studies in Nuclear Physics and Related Areas (ECT*)\\
and Fondazione Bruno Kessler, Strada delle Tabarelle 286, I-38123 Villazzano (TN), Italy}

\begin{abstract}
We make a systematic study of $\Lambda$ hyperon polarizations in unpolarized lepton induced semi-inclusive reactions such as $e^-N\to e^-\Lambda X$ and $e^+e^-\to\Lambda h X$. We present the general form of cross sections in terms of structure functions obtained from a general kinematic analysis. This already shows that the produced hyperons can be polarized in three orthogonal directions, i.e., the longitudinal direction along the hyperon momentum, the normal direction of the production plane, and the transverse direction in the production plane. We present the parton model results at the leading twist and leading order in perturbative QCD using transverse momentum dependent factorization and provide the expressions for these structure functions and polarizations in terms of three dimensional parton distribution functions and fragmentation functions. We emphasize in particular that by studying the longitudinal polarization and the transverse polarization in the production plane, we can extract the corresponding chiral-odd fragmentation functions $H_{1Lq}^{\perp\Lambda}$, $H_{1Tq}^{\Lambda}$ and $H_{1Tq}^{\perp\Lambda}$. We also present numerical results of rough estimates utilizing available parameterizations of fragmentation functions and approximations. We discuss how to measure these polarizations and point out in particular that they can be carried out in future EIC and/or $e^+e^-$ annihilation experiments such as Belle.
\end{abstract}

\maketitle

\section{Introduction}

Studies of the spin dependence of fragmentation functions (FFs) provide deep insight into the hadronization mechanism and important information on properties of the strong interaction. 
FFs are basic quantities to describe the fragmentation process of a high energy parton into hadrons. 
They are of non-perturbative nature and can be extracted from experimental data~\cite{Metz:2016swz}. 
Currently, such studies have been extended to the transverse momentum dependent (TMD) FFs. 
For the spin-1/2 hadron production, there are eight independent TMD FFs at the leading twist, and many more at higher twists. 
This is quite similar to that for TMD parton distribution functions (PDFs) of nucleon. 
However, while for TMD PDFs, there are already a few quantitative studies available~\cite{Zhang:2008nu,Lu:2009ip,Hautmann:2014kza,Angeles-Martinez:2015sea,Wei:2020glg,Echevarria:2020hpy,Bury:2020vhj,Bury:2021sue,Boglione:2021aha},  
TMD FFs are still quite poorly studied yet in experiments. 
Theoretical studies of leading and higher twist contributions to observables in semi-inclusive deep inelastic scattering (SIDIS) and $e^+e^-$ annihilation 
based on the collinear expansion can be found e.g. in Refs.~\cite{Ellis:1982wd, Ellis:1982cd,Qiu:1988dn,Qiu:1990xxa,Qiu:1990xy,Balitsky:1990ck,Levelt:1994np,Levelt:1993ac,Kotzinian:1994dv,Mulders:1995dh,Boer:1997mf,Kotzinian:1997wt,Boer:1997nt,Boer:1997qn,Bacchetta:2000jk,Boer:1999uu,Bacchetta:2004zf,Bacchetta:2006tn,Boer:2008fr,Eguchi:2006qz,Eguchi:2006mc,Koike:2006qv,Kanazawa:2013uia,Pitonyak:2013dsu,Yang:2016qsf,Liang:2006wp,Liang:2008vz,Gao:2010mj, Song:2010pf, Song:2013sja,Wei:2013csa,Wei:2014pma,Chen:2016moq,Wei:2016far,Yang:2017sxz}.

Lepton induced reactions such as SIDIS and $e^+e^-$ annihilation are described by the QCD factorization framework~\cite{Collins:1981uk,Collins:1981va,Collins:1984kg,Ji:2004wu}. 
They provide good places to study TMD FFs and have attracted much attentions both theoretically and experimentally.  
In this connection, the Belle collaboration has measured the transverse polarization of $\Lambda$ hyperons in $e^+e^-$ annihilation~\cite{Belle:2018ttu} and 
has inspired immediate parameterizations~\cite{DAlesio:2020wjq, Callos:2020qtu,Chen:2021hdn} and model calculations~\cite{Li:2020oto} of $D_{1Tq}^{\perp\Lambda}$. 
Theoretical studies have also been made to the transverse polarization of $\Lambda$ hyperon in SIDIS~\cite{Zhou:2008fb,Kang:2021ffh,Li:2021txj,Kang:2021kpt}. 
We note in particular that the parameterization in~\cite{Chen:2021hdn} emphasizes on the isospin symmetry 
while those in~\cite{DAlesio:2020wjq,Callos:2020qtu} include quite significant isospin symmetry violations. 
Because isospin symmetry is a fundamental property of strong interaction, it should be very important to make further tests using similar measurements in other processes. 

The Electron-Ion Collider (EIC) has been proposed and will be built as the next generation electron-proton collider. 
It will provide an excellent opportunity to study TMD FFs~\cite{Accardi:2012qut,Anderle:2021wcy,AbdulKhalek:2021gbh}. 
The SIDIS cross section at the leading twist is given by a convolution of TMD PDFs and FFs~\cite{Kotzinian:1994dv,Mulders:1995dh}. 
Even with the unpolarized nucleon, the number density $f_{1q}$ and the Boer-Mulders function $h_{1q}^\perp$ contribute. 
While $f_{1q}$ convolutes with $D_{1Tq}^{\perp}$ and gives rise to transverse hyperon polarization along the normal direction of the production plane, 
$h_{1q}^\perp$ couples with other chiral odd FFs and leads to the azimuthal asymmetry and longitudinal or transverse polarizations of final state hyperons. 
Therefore, rich physics can be studied in the future EIC experiment as long as proper observables are measured. 
The same situation applies to dihadron production in the $e^+e^-$ annihilation. 

In this paper, we present a phenomenological study on the three-dimensional polarization of $\Lambda$ hyperons produced in these reactions.  
We present the key results of the general kinematic analysis to obtain the general form of the differential cross section in terms of structure functions 
and those for these structure functions in the parton model at the leading twist and the leading order (LO) in perturbative QCD (pQCD). 
We discuss how to measure these polarizations in experiments and present a rough estimate of their magnitudes. 

The rest of this paper is organized as follows. 
In Sec.~II, we consider $e^-N\to e^- \Lambda X$, summarize the key formulae and present a rough estimate of the magnitudes of polarizations of $\Lambda$ using available parameterizations of FFs. 
In Sec.~III, we present the corresponding results for $e^+e^-\to \Lambda hX$.  
A short summary is given in Sec.~IV.

\section{$\Lambda$ polarizations in SIDIS} 
\label{sec:SIDIS}

In this section, we consider the unpolarized SIDIS $e^-N\to e^- \Lambda X$. 
We take the $\Lambda$ hyperon production as an explicit example. 
The obtained results can also apply to other spin-1/2 hadrons as well.  

For completeness, we start with a brief summary of the general kinematic analysis 
and present key results that are useful to guide future measurements and numerical estimates. 
We show the parton model results  at the leading twist and LO in pQCD. 
We discuss how to measure different components of $\Lambda$ polarization and present results of a rough estimate using 
parameterizations of FFs available in literature. 


\subsection{The kinematic analysis and $\Lambda$ polarizations}
\label{sec:SIDISkin}

\begin{figure}[htb]
\includegraphics[width=0.45\textwidth]{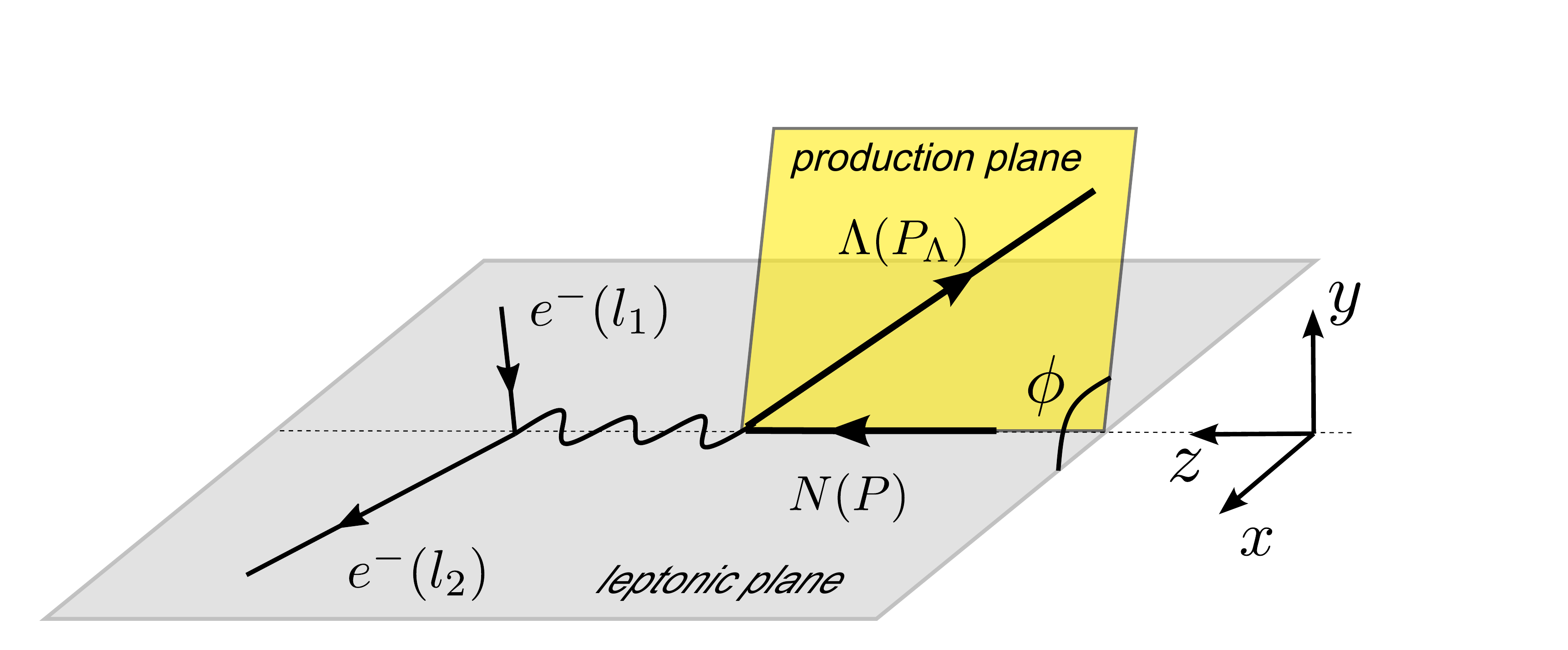}
\caption{Illustrating diagram of the coordinate system for $e^-N\to e^-\Lambda X$ in the photon-nucleon collinear frame. The symbol in the bracket denotes the four momentum of the particle.}
\label{fig:sidis-kinematics}
\end{figure}

We take the coordinate system for $e^-N\to e^-\Lambda X$ in the photon-nucleon collinear frame shown in Fig.~\ref{fig:sidis-kinematics}, 
where the incoming proton moves along $z$-direction, and the $x$-direction is determined by the transverse momenta of the leptons. 
The production plane of $\Lambda$ is determined by the momentum of the incoming proton and that of the outgoing $\Lambda$. 
The azimuthal angle $\phi$ is spanned by the transverse momentum of $\Lambda$ with respect to that of the outgoing electron.

The differential cross section is given by
\begin{align}
\frac{d\sigma^{\rm SIDIS}}{dxdydz_{\Lambda}d^2{P}_{\Lambda \perp}}
= \frac{ \pi \alpha_{\rm em}^2}{2Q^4} \frac{y}{z_\Lambda} L_{\mu\nu} W^{\mu\nu},
\label{eq:cs-sidis-ori}
\end{align}
where $Q^2=-q^2=xys$, $s=2l_1\cdot P$, $x=Q^2/2P\cdot q$, $y=P\cdot q/P\cdot l_1$, $z_\Lambda =P\cdot P_{\Lambda}/P\cdot q$,  
$P_{\Lambda\perp}=(0,0,\bm{P}_{\Lambda\perp})$ is the transverse momentum of $\Lambda$, 
$L_{\mu\nu}$ and $W^{\mu\nu}$ are the leptonic and hadronic tensors respectively. 

The hadronic tensor $W^{\mu\nu}$ is a function of $P$, $q$, $P_{\Lambda}$ and the spin polarization vector $S_\Lambda$ of $\Lambda$. 
It satisfies the general constraints imposed by the current conservation, the parity invariance and the hermiticity. 
In general, it is a sum of a set of ``basic Lorentz tensors'' $h_{\sigma i}^{\mu \nu}$ multiplied by corresponding scalar coefficients $W_i^\sigma$, i.e., 
\begin{align}
W^{S\mu\nu} = \sum_{i=1}^4 W_i^U h_{U i}^{S \mu \nu} + \sum_{i=1}^8 W_i^V h_{V i}^{S \mu \nu}, \label{eq:sidis-wmunu-kine}
\end{align}
where the first sum is over spin-independent terms and the second one is over spin-dependent ones. 
Here, we consider the unpolarized SIDIS, so that only the symmetric part $W^{S\mu\nu}$ (denoted by the superscript $S$) of the hadronic tensor contributes. 
The superscript $\sigma=U$ or $V$ of $W_i^\sigma$ or the subscript of $h_{\sigma i}^{\mu \nu}$ denotes the polarization of the final state $\Lambda$ hyperon. 

The complete set of $h_{\sigma i}^{\mu \nu}$ for the case of two spin-1/2 hadrons can be found in e.g. Refs.~\cite{Arnold:2008kf,Pitonyak:2013dsu,Chen:2016moq}. 
It has in particular been shown in~\cite{Chen:2016moq} that the spin dependent set can be constructed from the spin independent set 
by multiplying them with spin-dependent scalars and/or pseudo-scalars.
In this paper, we consider $e^-N\to e^-\Lambda X$ with unpolarized nucleon and need only the parity conserved symmetric ones. 
In this case, we obtain in total 12 independent basic Lorentz tensors. 
The complete unpolarized set is given by
\begin{align}
& h_{U i}^{S \mu \nu} =\left\{g^{\mu \nu}-\frac{q^{\mu} q^{\nu}}{q^{2}},~ P_{q}^{\mu} P_{q}^{\nu},~ P_{q}^{\{\mu} P_{\Lambda q}^{\nu\}},~ P_{\Lambda q}^{\mu} P_{\Lambda q}^{\nu}\right\}, 
\label{eq:bltU}
\end{align}
where $i=1\sim 4$, $a^{\{\mu} b^{\nu\}}\equiv a^\mu b^\nu + a^\nu b^\mu$,  and $a_q^\mu \equiv a^\mu-q^\mu(a\cdot q)/q^2$ satisfying $q\cdot a_q = 0$. 

The spin-dependent set is given by, 
 \begin{align}
& h_{V i}^{S \mu \nu}=\left\{\left[\lambda_\Lambda,~ (P_{\Lambda\perp} \cdot S_{\Lambda\perp})\right] \tilde{h}_{U j}^{S \mu \nu}, \varepsilon^{S_{\Lambda} q P P_{\Lambda}} h_{U k}^{S \mu \nu}\right\}, \label{eq:bltV}
\end{align}
where $\tilde{h}_{U j}^{S \mu \nu}=\big\{\varepsilon^{\{\mu q P P_{\Lambda}} P_{q}^{\nu\}},~ \varepsilon^{\{\mu q P P_{\Lambda}} P_{\Lambda q}^{\nu\}}\big\} (j=1,2)$, 
$\varepsilon^{\mu q P P_\Lambda} \equiv \varepsilon^{\mu\alpha\beta\rho} q_\alpha P_\beta P_{\Lambda\rho}$; 
$\lambda_\Lambda$ and $S_{\Lambda\perp}$ are the helicity and transverse components of the spin polarization vector of $\Lambda$;   
$i$ takes the values $1$ through $8$,  
and for $i=1,2$, $h_{V i}^{S \mu \nu}=\lambda_\Lambda \tilde{h}_{U j}^{S \mu \nu}$ $(j=1,2)$, corresponds to the longitudinal polarization of $\Lambda$, 
for $i=3,4$, $h_{V i}^{S \mu \nu}= (P_{\Lambda\perp} \cdot S_{\Lambda\perp})\tilde{h}_{U j}^{S \mu \nu}$ $(j=1,2)$, 
corresponds to the transverse polarization of $\Lambda$ in the production plane,  
and for $i=5\sim 8$,  $h_{V i}^{S \mu \nu}=\varepsilon^{S_{\Lambda} q P P_{\Lambda}} h_{U k}^{S \mu \nu}$ ($k=1 \sim 4$), 
corresponds to the transverse polarization of $\Lambda$ in the normal direction of the production plane.  
Totally, we have four spin independent and eight spin dependent basic Lorentz tensors.

We emphasize that the set of spin dependent tensor given by Eq.~(\ref{eq:bltV}) is complete but not unique. 
The advantages of using such a spin dependent set manifest themselves not only in the simple relationship between them and the unpolarized set 
but also in the differential cross section.  
We insert Eqs.~(\ref{eq:sidis-wmunu-kine}) through (\ref{eq:bltV}) into Eq.~(\ref{eq:cs-sidis-ori}) and obtain the general form of the differential cross section in terms of structure functions as, 
\begin{align}
& \frac{d\sigma^{\rm SIDIS}}{dxdydz_\Lambda d^2\bm{P}_{\Lambda\perp}} = \frac{2\pi  \alpha_{\rm em}^2}{xyQ^2} \Big\{ A(y) F_{UU}^T + B(y) F_{UU}^L \nonumber \\
& \phantom{X} + C(y) \cos\phi F_{UU}^{\cos\phi} + B(y) \cos2\phi F_{UU}^{\cos2\phi} \nonumber\\
& \phantom{X} + \lambda_\Lambda \left[ C(y) \sin\phi F_{UL}^{\sin\phi} + B(y) \sin2\phi F_{UL}^{\sin2\phi} \right] \nonumber\\
& \phantom{X} + S_{\Lambda T} \left[ C(y) \sin\phi F_{UT}^{\sin\phi} + B(y) \sin2\phi F_{UT}^{\sin2\phi} \right] \nonumber\\
& \phantom{X} + S_{\Lambda N} \Big[A(y) F_{UT}^T + B(y) F_{UT}^L + C(y) \cos\phi F_{UT}^{\cos\phi} \nonumber \\
& \phantom{XXXXX}  + B(y) \cos2\phi F_{UT}^{\cos2\phi}\Big] \Big\},
\label{eq:cs-sidis-kinematics}
\end{align}
where the $F$'s are structure functions that are Lorentz scalar functions of $(x, Q^2, z_\Lambda,{P}_{\Lambda\perp}^2)$, 
the superscript $T$ or $L$ specifies the polarization of the virtual photon and 
the two subscripts denote the polarization of nucleon and that of $\Lambda$ respectively; 
$S_{\Lambda N}$ is the component of the $\Lambda$ spin polarization vector in the normal direction of the production plane, 
and ${S}_{\Lambda T}$ is the transverse component in the production plane; 
the $y$ dependent coefficients are given by $A(y) = 1 + (1-y)^2$, $B(y) = 2(1-y)$ and $C(y) = 2(2-y)\sqrt{1-y}$. 

From Eq.~(\ref{eq:cs-sidis-kinematics}), we see clearly that due to the symmetric form taken by the spin dependent Lorentz tensors 
given by Eq.~(\ref{eq:bltV}), there are also similar symmetries in form for the $\Lambda$ spin dependent cross sections. 
In particular, the $S_{\Lambda N}$ dependent cross section has the same structure with the unpolarized part and the $\lambda_\Lambda$ and $S_{\Lambda T}$ dependent parts take the same structure. 
Another reason for us to write Eq.~(\ref{eq:cs-sidis-kinematics}) in this form is that we can easily calculate the $\Lambda$ polarizations along these three directions 
by calculating the average values of $\lambda_\Lambda$, $S_{\Lambda T}$ and $S_{\Lambda N}$ respectively~\cite{Chen:2016moq}. 
We also note that if we consider SIDIS with polarized nucleon beam and sum over the spin of $\Lambda$, we can obtain the differential cross section 
in exactly the same form as that given by Eq.~(\ref{eq:cs-sidis-kinematics}) with replacements of the polarization vector $S_\Lambda$ by that of the nucleon 
and the corresponding structure functions $F_{Uj}$ by $F_{jU}$ where in the latter case $j=U, L, T$ or $N$ denotes polarization of the nucleon. 
However, we should also note that in this case longitudinal $L$ denotes the direction of momentum of the incoming nucleon while the transverse directions $T$ and $N$ 
are determined by momentum of the nucleon and that of $\Lambda$ and they are different in different events. 

From Eq. (\ref{eq:cs-sidis-kinematics}), we also see immediately that although the colliding leptons and nucleons are unpolarized, the produced $\Lambda$ hyperons are polarized in three directions, i.e., the longitudinal direction, the normal direction of the production plane and the transverse direction inside the production plane. To be more specific, at the given azimuthal angle $\phi$, different components of the $\Lambda$ polarization are expressed by the corresponding structure functions. They are given by 
\begin{align}
&\mathcal{P}_L= \frac{1}{{\cal F}_{UU}^{\rm tot}} 
\Big\{C(y) \sin\phi F_{UL}^{\sin\phi} + B(y) \sin2\phi F_{UL}^{\sin2\phi} \Big\}, 
\label{eq:PL-SIDIS-kin} \\
&{\cal P}_T 
= \frac{1}{{\cal F}_{UU}^{\rm tot}}  
\Big\{C(y) \sin\phi F_{UT}^{\sin\phi} + B(y) \sin2\phi F_{UT}^{\sin2\phi}\Big\}, 
\label{eq:PT-SIDIS-kin}\\
&{\cal P}_N 
= \frac{1}{{\cal F}_{UU}^{\rm tot}} \Big\{A(y) F_{UT}^T + B(y) F_{UT}^L  \nonumber\\ 
&~~~~~~~~~ + C(y) \cos\phi F_{UT}^{\cos\phi}+ B(y) \cos2\phi F_{UT}^{\cos2\phi} \Big\}, 
\label{eq:PN-SIDIS-kin}
\end{align}
where ${\cal F}_{UU}^{\rm tot}$ is given by, 
\begin{align}
{\cal F}_{UU}^{\rm tot} = & A(y) F_{UU}^T + B(y) F_{UU}^L + C(y) \cos\phi F_{UU}^{\cos\phi}+ \nonumber \\ 
& + B(y) \cos2\phi F_{UU}^{\cos2\phi}, \label{eq:Ftot}
\end{align}
and is related directly to the unpolarized cross section as 
\begin{align}
&\frac{d\sigma_{UU}^{\rm SIDIS}} {dxdydz_\Lambda d^2 \bm{P}_{\Lambda\perp}} = \frac{2\pi \alpha_{\rm em}^2}{xyQ^2} {\cal F}_{UU}^{\rm tot}.\label{xsection-sidis-summed}
\end{align}
Here, we use $\mathcal{P}_j$ ($j=L, T, N$) to denote different components of the polarization from the un-integrated differential cross section given by Eq.~(\ref{eq:cs-sidis-kinematics}).
They are functions of $x,y,z_\Lambda, {P}_{\Lambda\perp}^2$ and $\phi$.  

We note in particular from Eqs.~(\ref{eq:PL-SIDIS-kin})-(\ref{eq:PN-SIDIS-kin}) that all the three components $\mathcal{P}_L$, $\mathcal{P}_T$ and ${\cal P}_N$ 
of the polarization of $\Lambda$ are azimuthal angle $\phi$ dependent. 
If we integrate over $\phi$, the cross section is reduced to
\begin{align}
& \frac{d\sigma^{\rm SIDIS}}{dxdydz_\Lambda d {P}_{\Lambda\perp}^2} = \frac{2\pi^2 \alpha_{\rm em}^2}{xyQ^2} 
\Big\{ A(y) F_{UU}^T + B(y) F_{UU}^L  \nonumber \\
& ~~~~~~~~ + S_{\Lambda N} \Big[A(y) F_{UT}^T + B(y) F_{UT}^L \Big] \Big\},
\label{eq:cs-sidis-kinematics2}
\end{align}
and one finds that the polarizations become
\begin{align}
& \langle {\cal P}_L \rangle =\langle {\cal P}_T \rangle= 0, \label{eq:PLT-SIDIS-kin2} \\
& \langle {\cal P}_N \rangle= \frac {A(y) F_{UT}^T + B(y) F_{UT}^L } {A(y) F_{UU}^T + B(y) F_{UU}^L }. \label{eq:PN-SIDIS-kin2} 
\end{align}
Here as well as in the following of this paper, we use $\langle {\cal P}_j \rangle$ to denote different components of the $\Lambda$ polarization integrated over $\phi$.   
They are functions of $x,y,z_\Lambda$ and ${P}_{\Lambda\perp}^2$.  
We see that both ${\cal P}_L$ and ${\cal P}_T$ vanish after integrating over $\phi$ in the whole phase space,  
and the only survival component is ${\cal P}_N$ that is in the normal direction of the production plane.  
In this case, we can only study the spin-dependent but azimuthal angle independent structure functions $F_{UT}^T$ and $F_{UT}^L$. 

To study those both spin and azimuthal angle dependent structure functions involved in ${\cal P}_L$, ${\cal P}_T$ and ${\cal P}_N$, 
we cannot simply carry out the whole phase space average over the azimuthal angle. 
For this purpose, we propose two ways to measure these polarizations in experiments 
and describe them in details in the following. 

\subsubsection{Polarizations in given quadrants}

The first way is to divide the $2\pi$ phase space of the azimuthal angle, as usual, into four quadrants, 
take the combination of two of them and take the average respectively. 
More precisely, we obtain
\begin{align}
& \langle \mathcal{P}_L\rangle _{\rm I+II}=-\langle \mathcal{P}_L\rangle _{\rm III+IV} 
  =\frac{2}{\pi} 
\frac{C(y) F_{UL}^{\sin\phi}}{A(y) F_{UU}^T + B(y) F_{UU}^L},
\label{eq:PL-SIDIS-0-pi}\\
& \langle \mathcal{P}_L\rangle _{\rm I+III}=-\langle \mathcal{P}_L\rangle _{\rm II+IV} 
  =\frac{2}{\pi}\frac{B(y) F_{UL}^{\sin2\phi}}{ A(y) F_{UU}^T + B(y) F_{UU}^L },
\label{eq:PL-SIDIS-0-pi2}\\
& \langle \mathcal{P}_T\rangle _{\rm I+II}=-\langle \mathcal{P}_T\rangle _{\rm III+IV} 
  =\frac{2}{\pi} 
\frac{C(y) F_{UT}^{\sin\phi}}{A(y) F_{UU}^T + B(y) F_{UU}^L},
\label{eq:PT-SIDIS-0-pi}\\
&  \langle \mathcal{P}_T\rangle _{\rm I+III}=-\langle \mathcal{P}_T\rangle _{\rm II+IV} 
  =\frac{2}{\pi}\frac{B(y) F_{UT}^{\sin2\phi}}{ A(y) F_{UU}^T + B(y) F_{UU}^L },
                              \label{eq:PT-SIDIS-0-pi2}\\
& \langle \mathcal{P}_N\rangle_{\rm I+IV} =\frac{A(y)F_{UT}^T+B(y)F_{UT}^L+\frac{2}{\pi} C(y)F_{UT}^{\cos\phi}}{A(y)F_{UU}^T+B(y)F_{UU}^L+\frac{2}{\pi} C(y)F_{UU}^{\cos\phi}},\\
& \langle \mathcal{P}_N\rangle_{\rm II+III} =\frac{A(y)F_{UT}^T+B(y)F_{UT}^L -\frac{2}{\pi}C(y)F_{UT}^{\cos\phi}}{A(y)F_{UU}^T+B(y)F_{UU}^L-\frac{2}{\pi}C(y)F_{UU}^{\cos\phi}}.
    \label{eq:PN-SIDIS-0-pi}
\end{align}
Here, I$\sim$ IV denotes different quadrants; the subscript $\rm I+II$ denotes to select events where $\Lambda$ hyperons are in the first and second quadrants, and the same applies to others. By limiting the phase space of the azimuthal angle, we pick up the corresponding azimuthal angle dependent structure functions. 

It is also clear that, to study $F_{UT}^{\cos2\phi}$, we need to switch the integration region from I$\sim$IV to i$\sim$iv with an anticlockwise rotation of $\pi/4$. 
The new quadrants i$\sim$iv correspond to $(-\pi/4, \pi/4)$, $(\pi/4,3\pi/4)$, $(3\pi/4, 5\pi/4)$ and $(5\pi/4,7\pi/4)$. Therefore, we have
\begin{align}
& \langle \mathcal{P}_N\rangle_{\rm i+iii} =\frac{A(y)F_{UT}^T+B(y)F_{UT}^L+\frac{2}{\pi}B(y)F_{UT}^{\cos2\phi}}{A(y)F_{UU}^T+B(y)F_{UU}^L+\frac{2}{\pi}B(y)F_{UU}^{\cos2\phi}}, \\
& \langle \mathcal{P}_N\rangle_{\rm ii+iv} =\frac{A(y)F_{UT}^T+B(y)F_{UT}^L-\frac{2}{\pi}B(y)F_{UT}^{\cos2\phi}}{A(y)F_{UU}^T+B(y)F_{UU}^L-\frac{2}{\pi}B(y)F_{UU}^{\cos2\phi}}. \label{eq:PN-SIDIS-0-pi2} 
\end{align}

\subsubsection{The $\sin n\phi$ or $\cos n\phi$ weighted polarizations}

The second way is to measure the polarizations weighted by $\sin n\phi$ or $\cos n\phi$. 
More precisely, we take the average 
\begin{align}
& \langle \mathcal{P}_{j}^{\sin n\phi}\rangle \equiv 
\frac{\int d\phi \frac{d\sigma_{UU}^{\rm SIDIS}}{dxdydz_\Lambda d^2 {P}_{\Lambda\perp}} \mathcal{P}_{j} (x,y,z_\Lambda,\bm{P}_{\Lambda\perp}) \sin n\phi}
{\int d\phi \frac{d\sigma_{UU}^{\rm SIDIS}}{dxdydz_\Lambda d^2{P}_{\Lambda\perp}}},\label{eq:PLTN:weighted-1}
\\
& \langle \mathcal{P}_{j}^{\cos n\phi}\rangle \equiv 
\frac{\int d\phi \frac{d\sigma_{UU}^{\rm SIDIS}}{dxdydz_\Lambda d^2{P}_{\Lambda\perp}} \mathcal{P}_{j} (x,y,z_\Lambda,\bm{P}_{\Lambda\perp}) \cos n\phi}
{\int d\phi \frac{d\sigma_{UU}^{\rm SIDIS}}{dxdydz_\Lambda d^2{P}_{\Lambda\perp}}},\label{eq:PLTN:weighted-2}
\end{align}
where $j=L,T,N$ denotes different components of $\mathcal{P}$ and $n=1,2$. 
It is straightforward to obtain
\begin{align}
&\langle \mathcal{P}_L^{\sin\phi}\rangle = \frac{1}{2} \frac{C(y) F_{UL}^{\sin\phi}}{A(y) F_{UU}^T + B(y) F_{UU}^L}, \\
&\langle \mathcal{P}_L^{\sin2\phi}\rangle= \frac{1}{2} \frac{B(y) F_{UL}^{\sin2\phi}}{A(y) F_{UU}^T + B(y) F_{UU}^L}, \\
&\langle \mathcal{P}_T^{\sin\phi}\rangle = \frac{1}{2} \frac{C(y) F_{UT}^{\sin\phi}}{A(y) F_{UU}^T + B(y) F_{UU}^L}, \\
&\langle \mathcal{P}_T^{\sin2\phi}\rangle = \frac{1}{2} \frac{B(y) F_{UT}^{\sin2\phi}}{A(y) F_{UU}^T + B(y) F_{UU}^L},\\
&\langle \mathcal{P}_N^{\cos\phi}\rangle = \frac{1}{2} \frac{C(y) F_{UT}^{\cos\phi}}{A(y) F_{UU}^T + B(y) F_{UU}^L}, \\
&\langle \mathcal{P}_N^{\cos2\phi}\rangle = \frac{1}{2} \frac{B(y) F_{UT}^{\cos2\phi}}{A(y) F_{UU}^T + B(y) F_{UU}^L}.
\end{align}

We see that such $\sin n\phi$ or $\cos n\phi$ weighted polarizations are equivalent to the polarizations in the given quadrants, except for the difference in the overall normalization. 
In the following of this paper, we only present results for these $\sin n\phi$ or $\cos n\phi$ weighted polarizations. 
The polarizations in the given quadrants can be obtained accordingly.

We emphasize that the $\sin n\phi$ or $\cos n\phi$ weighted polarizations can also be directly measured in experiments. 
We recall that the $\Lambda$ polarization can be measured from $\langle\cos\theta^*\rangle$ 
where $\theta^*$ is the angle between the momentum direction of the decay product $p$ in $\Lambda\to p\pi^-$ and the polarization direction in the $\Lambda$ rest frame. 
These $\sin n\phi$ or $\cos n\phi$ weighted polarizations can be simply obtained by measuring the corresponding $\langle\sin n\phi \cos\theta^*\rangle$ or $\langle\cos n\phi \cos\theta^*\rangle$. 
We present the precise formulae in appendix~\ref{sec:app:measurement}. 

\subsection{Parton model results}

At the LO in pQCD and leading twist, the parton model results can be found in different literatures e.g.~\cite{Mulders:1995dh,Yang:2016qsf}. We summarize the key results here. The hadronic tensor is given by
\begin{align}
W^{\mu\nu} = ~ & 2z_\Lambda e_q^2 \int d^2 \bm{k}_\perp d^2 \bm{p}_T 
\delta^2 (z_\Lambda \bm{k}_\perp + \bm{p}_T - \bm{P}_{\Lambda\perp}) \nonumber\\
&\times{\rm Tr} \left[
2\hat\Phi_q (x, \bm{k}_\perp) \gamma^\mu 2 \hat\Xi_q ^\Lambda (z_\Lambda,\bm{p}_T) \gamma^\nu
\right],\label{wmunu-sidis}
\end{align}
where $\bm{k}_\perp$ is the transverse momentum of the struck quark with respect to the proton momentum, 
$\bm{p}_T$ is the transverse momentum of the produced $\Lambda$ hyperon with respect to the direction of the fragmenting quark, and, 
$\hat\Phi_q(x, \bm{k}_\perp)$ and $\hat\Xi_q^\Lambda(z_\Lambda,\bm{p}_T)$ are quark-quark correlators. 
A sum over quark and anti-quark flavor $q$ is implicit.
For the unpolarized nucleon and polarized $\Lambda$ hyperon, we have 
\begin{align}
&4 \hat\Phi_q (x, {k}_\perp) 
= \slashed n_+ f_{1q}  + \frac{i [\slashed k_\perp, \slashed n_+]}{2m_p} h_{1q}^\perp , 
\label{correlation-matrix-phi}\\
&4 \hat\Xi_q^\Lambda (z_\Lambda,{p}_T)= \slashed n_- \Bigl[
D_{1q}^{\Lambda}  + \frac{(\bm{\widehat e}_j \times \bm{p}_T) \cdot \bm{S}_{\Lambda\perp}}{z_\Lambda M_\Lambda}  D_{1Tq}^{\perp\Lambda} 
\Bigr]\nonumber\\
  &\phantom{XXX} + \gamma_5 \slashed n_- \Bigl[
\lambda_\Lambda G_{1Lq}^\Lambda 
+ \frac{p_T \cdot S_{\Lambda\perp}}{z_\Lambda M_\Lambda} G_{1Tq}^{\perp\Lambda} 
\Bigr]\nonumber\\
&\phantom{XXX}+ \frac{i  [\slashed p_T, \slashed n_-] }{2M_\Lambda} H_{1q}^{\perp\Lambda} 
+ \frac{1}{2} [\slashed S_{\Lambda\perp}, \slashed n_-] \gamma_5 H_{1Tq}^{\Lambda} \nonumber\\
&\phantom{XXX} + \frac{[\slashed p_T, \slashed n_-] \gamma_5 }{2M_\Lambda} \Bigl[\lambda_\Lambda H_{1Lq}^{\perp\Lambda} 
    + \frac{p_T \cdot S_{\Lambda\perp}}{M_\Lambda} H_{1Tq}^{\perp\Lambda}  \Bigr],  \label{correlation-matrix-Xi}
\end{align}
where $[\slashed a, \slashed b] \equiv \slashed a \slashed b - \slashed b \slashed a$, $\bm{\widehat e}_j$ 
denotes the direction of the fragmenting quark, $n_+$ and $n_-$ are unit vectors in the light-cone coordinate. 

By inserting Eqs.~(\ref{correlation-matrix-phi}) and (\ref{correlation-matrix-Xi}) into Eq.~(\ref{wmunu-sidis}), 
carrying out traces, and making Lorentz contractions with the leptonic tensor, we obtain expressions of the structure functions at the leading twist. 
They are given by
\begin{align}
& F_{UU}^T = {\cal I} \left[ f_1  D_1^\Lambda \right],\label{eq:1:FUUT} \\
& F_{UU}^{\cos2\phi} = {\cal I} \left[ w_2  h_{1}^\perp  H_{1}^{\perp\Lambda} \right],\label{eq:1:FUUcos2phi}\\
& F_{UL}^{\sin2\phi} = {\cal I} \left[  w_2  h_{1}^\perp  H_{1L}^{\perp\Lambda} \right],\label{eq:1:FUUsin2phi}\\
& F_{UT}^T = {\cal I} \left[  \bar w_1 f_{1}  D_{1T}^{\perp\Lambda} \right]/z_\Lambda,\label{eq:1:FUTT}\\
& F_{UT}^{\cos2\phi} = {\cal I}[ w_1 h_1^\perp H_{1T}^{\Lambda}]+{\cal I} [ w_{3b} h_{1}^\perp   H_{1T}^{\perp\Lambda} ], \label{eq:1:FUTcos2phi}\\
& F_{UT}^{\sin2\phi} = -{\cal I}[ w_1 h_1^\perp H_{1T}^{\Lambda}] + {\cal I} [w_{3a} h_{1}^\perp   H_{1T}^{\perp\Lambda}], \label{eq:1:FUTsin2phi}\\
&F_{UU/UT}^L=F_{UU/UT}^{\cos\phi}=F_{UL/UT}^{\sin\phi}=0. \label{eq:1:Fzeros}
\end{align}
Here ${\cal I}[wfD]$ denotes the convolution defined as ${\cal I}[wfD] \equiv e_q^2 x\int  d^2{p}_T d^2{k}_\perp \delta^2 (z_\Lambda {k}_\perp +{p}_T - {P}_{\Lambda\perp}) w f_q (x, k_\perp) D_q (z_\Lambda, p_T)$ and
\begin{align}
& w_1 = -{\hat P_{\Lambda\perp} \cdot k_\perp}/{m_p},  ~~~~~ \bar w_1 = -{\hat P_{\Lambda\perp} \cdot p_T}/{M_\Lambda},  \label{eq:wi-1}\\
& w_2 = -\Bigl[{2(\hat P_{\Lambda\perp}\cdot k_\perp)(\hat P_{\Lambda\perp}\cdot p_T) + k_\perp\cdot p_T}\Bigr]/{m_p M_\Lambda}, \label{eq:wi-3}\\
& w_{3a} = -\Big[ 2(\hat P_{\Lambda\perp}\cdot k_\perp) (\hat P_{\Lambda\perp}\cdot p_T)^2 \nonumber\\
&\phantom{XXXX}+( k_\perp\cdot p_T)(\hat P_{\Lambda\perp}\cdot p_T)\Big] /{m_p M_\Lambda^2}, \label{eq:wi-4A}\\
& w_{3b} =  - \Big[2(\hat P_{\Lambda\perp}\cdot k_\perp)(\hat P_{\Lambda\perp}\cdot p_T)^2 +( k_\perp\cdot p_T)(\hat P_{\Lambda\perp}\cdot p_T) \nonumber\\
&\phantom{XXXX} + (\hat P_{\Lambda\perp} \cdot k_\perp)p_T^2 \Big]/{m_p M_\Lambda^2}, \label{eq:wi-4B}
\end{align}
where $\hat P_{\Lambda\perp}=P_{\Lambda\perp}/|\bm{P}_{\Lambda\perp}|$ is the unit vector along the direction of the transverse momentum of $\Lambda$. 

By inserting Eqs. (\ref{eq:1:FUUT})-(\ref{eq:1:Fzeros}) into Eqs.~(\ref{eq:PL-SIDIS-kin})-(\ref{eq:PN-SIDIS-kin}), we obtain the expressions of $\Lambda$ polarizations in $e^-N\to e^-\Lambda X$ 
in terms of PDFs and FFs. 
The $\phi$-integrated transverse polarization in the normal direction and the $\sin n\phi$ and $\cos n\phi$ weighted polarizations are given by,
\begin{align} 
&\langle {\cal P}_N\rangle =\frac{1}{z_\Lambda}\frac{{\cal I}[\bar w_1 f_1 D_{1T}^{\perp\Lambda}]}{{\cal I}[f_1 D_1^\Lambda]}, \label{eq:PNaveSIDISLT}\\
&\langle {\cal P}_L^{\sin\phi}\rangle=\langle {\cal P}_T^{\sin\phi}\rangle =\langle {\cal P}_N^{\cos\phi}\rangle =0, \label{eq:PsinphiSIDISLT} \\
&\langle {\cal P}_L^{\sin2\phi}\rangle=\frac{B(y)}{2A(y)} \frac{{\cal I}[w_2h_1^\perp H_{1L}^{\perp\Lambda}]}{{\cal I}[f_1D_1^\Lambda]}, \label{eq:PLsin2phiSIDISLT}\\
&\langle {\cal P}_T^{\sin2\phi}\rangle=\frac{B(y)}{2A(y)} \frac{- {\cal I}[ w_1 h_1^\perp H_{1T}^{\Lambda}]+{\cal I} [w_{3a} h_{1}^\perp   H_{1T}^{\perp\Lambda}] }{{\cal I}[f_1D_1^\Lambda]}, \label{eq:PTsin2phiSIDISLT}\\
&\langle {\cal P}_N^{\cos2\phi}\rangle=\frac{B(y)}{2A(y)} \frac{{\cal I}[w_1 h_1^\perp H_{1T}^\Lambda]+{\cal I}[w_{3b}h_1^\perp H_{1T}^{\perp\Lambda}]}{{\cal I}[f_1D_1^\Lambda]}.  \label{eq:PNcos2phiSIDISLT}
\end{align}

We see that at the leading twist $\langle {\cal P}_N\rangle$ is a convolution of $f_1$ with $D_{1T}^{\perp\Lambda}$, 
the other three components such as $\langle {\cal P}_L^{\sin2\phi}\rangle$,  $\langle {\cal P}_T^{\sin2\phi}\rangle$ and $\langle {\cal P}_N^{\cos2\phi}\rangle$ are convolutions of 
the chiral odd PDF $h_1^\perp$ with the corresponding chiral odd spin dependent FFs. 
Since $h_1^\perp$  is expected to be smaller than $f_1$,  the magnitudes of these three components are expected to be smaller than that of $\langle {\cal P}_N\rangle$ in general. 
On the other hand, the corresponding spin dependent FFs are interesting and it is worthwhile to do such measurements in future EIC with high luminosities. We make a rough estimate of their magnitudes in the next subsection.

\subsection{A rough numerical estimate}

Having the parton model results of different components of $\Lambda$ polarizations in terms of FFs and PDFs, 
we can calculate them if the corresponding FFs and PDFs are given. 
Currently, only parameterizations of a few of them with limited accuracy are available so we present only a rough estimate in the following of this section. 

As usual, we take the Gaussian ansatz for the transverse momentum dependence of PDFs and/or FFs, i.e., 
\begin{align}
& f_{1q} (x, k_\perp) = f_{1q} (x) \frac{1}{\pi \Delta_{p}^2} e^{ - \bm{k}_\perp^2/\Delta^2_{p}} , \\
& D_{1q}^\Lambda (z, p_T) = D_{1q}^\Lambda (z) \frac{1}{\pi \Delta_{\Lambda}^2} e^{-\bm{p}_T^2/\Delta^2_{\Lambda}}, \label{eq:unpolFFGaussian}
\end{align}
where $\Delta_p$ and $\Delta_\Lambda$ are the corresponding Gaussian widths.  
Currently, this ansatz is commonly used~\cite{DAlesio:2020wjq,Callos:2020qtu} and the widths are determined approximately as $\Delta_p^2=0.61$ GeV$^2$ and $\Delta_{\Lambda}^2 = 0.118$ GeV$^2$.

For the polarized TMD FFs, we take the same factorized form of $z$ and $p_T$ dependences and similar Gaussian form of the $p_T$ dependence. 
Since there is almost no experimental data available yet,  
we simply take the Gaussian widths of the polarized FFs as the same with that of the unpolarized FF to minimize free parameters to make a rough estimate here. 

We emphasize that the $p_T$ dependences of TMD PDFs and/or FFs may in general deviate from the simple Gaussian form, especially after taking into account 
the TMD evolution determined by the Collins-Soper evolution equation~\cite{Collins:1981uk,Collins:1981va,Collins:1984kg}. 
While the TMD evolution can be calculated~\cite{Idilbi:2004vb}, 
the exact form of the $p_T$ dependence and the free parameters at the initial scale can only be extracted from the experimental data. 
Since there is still no experimental data available yet for the chiral-odd FFs involved for calculating different components of $\Lambda$ polarizations, 
we just take such simple Gaussian ansatz for a rough estimate. 

Using such Gaussian ansatz, we obtain the $|\bm{P}_{\Lambda\perp}|$-integrated $\Lambda$ polarizations.   
The $\phi$ averaged transverse polarization ${\cal P}_N$ and the $\sin 2\phi$ and $\cos 2\phi$ weighted polarizations are given by
\begin{align}
&\langle\bar{\mathcal{P}}_N \rangle 
=\frac{\sqrt{\pi} \kappa_3(z_\Lambda)}{2z_\Lambda} \frac{e_q^2 xf_{1q} (x) D_{1Tq}^{\perp\Lambda} (z_\Lambda)}{e_q^2 xf_{1q}(x) D_{1q}^{\Lambda} (z_\Lambda)}, \label{eq:PN-SIDIS}\\
&\langle \bar{ {\cal P}}_L^{\sin 2\phi} \rangle 
=- \kappa_1(z_\Lambda) \frac{B(y)}{2A(y)} \frac{e_q^2 h_{1q}^\perp (x) H_{1Lq}^{\perp\Lambda} (z_\Lambda) }{e_q^2 f_{1q}(x) D_{1q}^\Lambda (z_\Lambda)},\label{eq:PLsin2phi-SIDIS}\\
&\langle \bar {\cal P}_T^{\sin 2\phi} \rangle
=\sqrt{\pi}\kappa_2(z_\Lambda) \frac{B(y)}{4A(y)}  \nonumber\\
&\phantom{XX}\times \frac{e_q^2 h_{1q}^\perp (x) \left[-H_{1Tq}^\Lambda (z_\Lambda) + \kappa_4(z_\Lambda)H_{1Tq}^{\perp\Lambda} (z_\Lambda)\right] } {e_q^2 f_{1q} (x) D_{1q}^\Lambda (z_\Lambda)}, \label{eq:PTsin2phi-SIDIS}\\
&\langle \bar{ {\cal P}}_N^{\cos 2\phi} \rangle =\sqrt{\pi}\kappa_2(z_\Lambda) \frac{B(y)} {4A(y)} \nonumber\\
& \phantom{XX} \times \frac{e_q^2 h_{1q}^\perp(x) \left[ H_{1Tq}^\Lambda(z_\Lambda) + \kappa_5(z_\Lambda) H_{1Tq}^{\perp\Lambda}(z_\Lambda)  \right]} {e_q^2 f_{1q}(x) D_{1q}^\Lambda(z_\Lambda)}, \label{eq:PNcos2phi-SIDIS}
\end{align}
where $\langle\bar{\mathcal{P}}_j \rangle$ denotes different components of the $\Lambda$ polarization after integrating over $d^2 P_{\perp\Lambda}$ 
and they are functions of $(x,y,z_\Lambda)$;   $\kappa_i$'s depend on $z_\Lambda$ and the Gaussian widths,  
\begin{align}
&\kappa_1(z_\Lambda)= {z_\Lambda \Delta_p^2 \Delta_\Lambda^2 }/{m_p M_\Lambda(\Delta_\Lambda^2 + z_\Lambda^2 \Delta_p^2 )}, \label{eq:kappa1} \\
&\kappa_2(z_\Lambda) ={z_\Lambda\Delta^2_{p} }/{m_p\sqrt{\Delta^2_{\Lambda}+z_\Lambda^2\Delta^2_{p}}}, \\ 
&\kappa_3(z_\Lambda) = {\Delta_\Lambda^2}/{M_\Lambda \sqrt{ \Delta_\Lambda^2 + z_\Lambda^2 \Delta_p^2 }}, \\ 
&\kappa_4(z_\Lambda) ={\Delta^2_{\Lambda} (2\Delta^2_{\Lambda} + z_\Lambda^2\Delta^2_{p})}/{2M_\Lambda^2(\Delta^2_{\Lambda} + z_\Lambda^2 \Delta^2_{p})}, \\
&\kappa_5(z_\Lambda)={\Delta^2_{\Lambda} (\Delta^2_{\Lambda} - z_\Lambda^2\Delta^2_{p})}/{2M_\Lambda^2(\Delta^2_{\Lambda} + z_\Lambda^2 \Delta^2_{p})}. \label{eq:kappa5}
\end{align}

\begin{figure}[h!]
\includegraphics[width=0.4\textwidth]{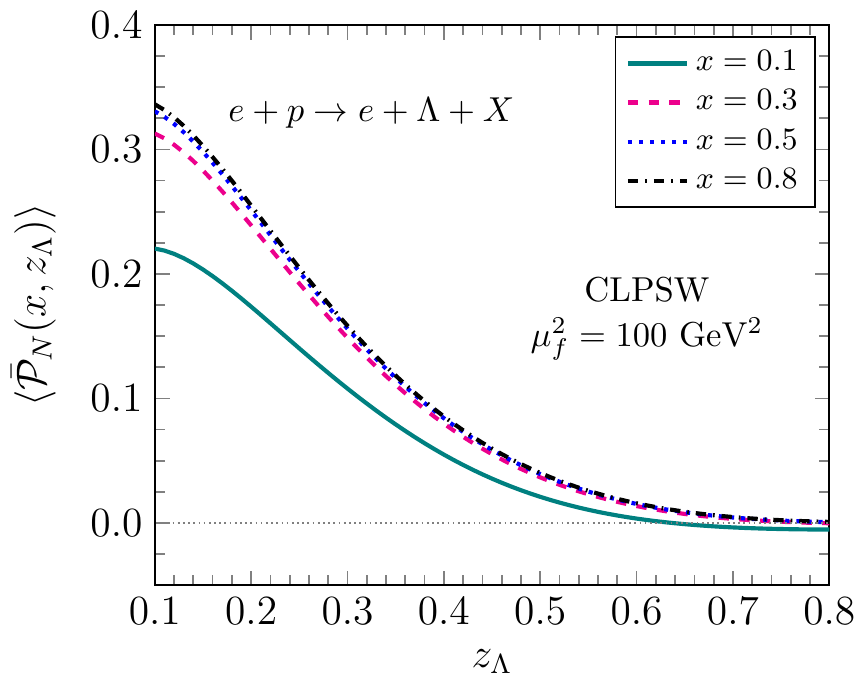}
\caption{Calculated results for the transverse polarization $\langle\bar{\cal P}_N\rangle$ in $e^-N\to e^-\Lambda X$ as functions of $z_\Lambda$ at different values of $x$ using the CLPSW parametrization~\cite{Chen:2021hdn} of $D_{1T}^{\perp\Lambda}$. 
The unpolarized PDFs and FFs are taken from NLO CT14~\cite{Dulat:2015mca} and DSV~\cite{deFlorian:1997zj}.}
\label{fig:sidis-tpol-1}
\end{figure}

With Eq.~(\ref{eq:PN-SIDIS}), we estimate the magnitude of the $\Lambda$ polarization in SIDIS along the normal direction of the production plane 
using the parameterizations of $D_{1T}^{\perp\Lambda}$ available. 
The universality of the polarized FF $D_{1T}^{\perp\Lambda}$ has been proven~\cite{Metz:2002iz,Meissner:2008yf,Boer:2010ya}. 
The results obtained by using the Chen-Liang-Pan-Song-Wei (CLPSW) parametrization~\cite{Chen:2021hdn} are given by Fig.~\ref{fig:sidis-tpol-1}. 
Here, we chose a typical value of $Q=10~{\rm GeV}$ and set the factorization scale $\mu_f=Q$. 
The results are obtained for different values of $x$ covered by the kinematic range of the future EIC or EicC~\cite{Anderle:2021wcy,AbdulKhalek:2021gbh}.

From Fig.~\ref{fig:sidis-tpol-1}, we see that the magnitude of $\langle\bar{\cal P}_N\rangle$ can reach as large as $30\%$ at small $z_\Lambda$. 
We also see that $\langle\bar{\cal P}_N\rangle$ increases with increasing $x$ and reaches a limit at large $x$. 
This is because at large $x$ the valence quarks $u$ and $d$ dominate and $D_{1Tu}^{\perp\Lambda}$ and $D_{1Td}^{\perp\Lambda}$ are equal due to the isospin symmetry.  

We recall that there are two other parameterizations of $D_{1T}^{\perp\Lambda}$ available, i.e., the D'Alesio-Murgia-Zaccheddu (DMZ)~\cite{DAlesio:2020wjq} 
and the Callos-Kang-Terry (CKT)~\cite{Callos:2020qtu} parameterizations.
Though both of them describe the Belle data~\cite{Belle:2018ttu}, there is a distinct difference between them and the CLPSW parameterization~\cite{Chen:2021hdn}. 
Significant isospin symmetry violations in $D_{1T}^{\perp\Lambda}$ are included in these two parameterizations~\cite{DAlesio:2020wjq,Callos:2020qtu} 
while it is conserved in the CLPSW parametrization~\cite{Chen:2021hdn}.  
Although it is very unlikely that such significant isospin symmetry violations exist in FFs, it is important to test it in experiments.  
In EIC, one can change the target nucleon and/or nucleus so that contributions from $u$ and $d$-quarks change drastically. 
Therefore, it provides an ideal place to test the isospin symmetry of polarized FFs.

\begin{figure}[h!]
\includegraphics[width=0.4\textwidth]{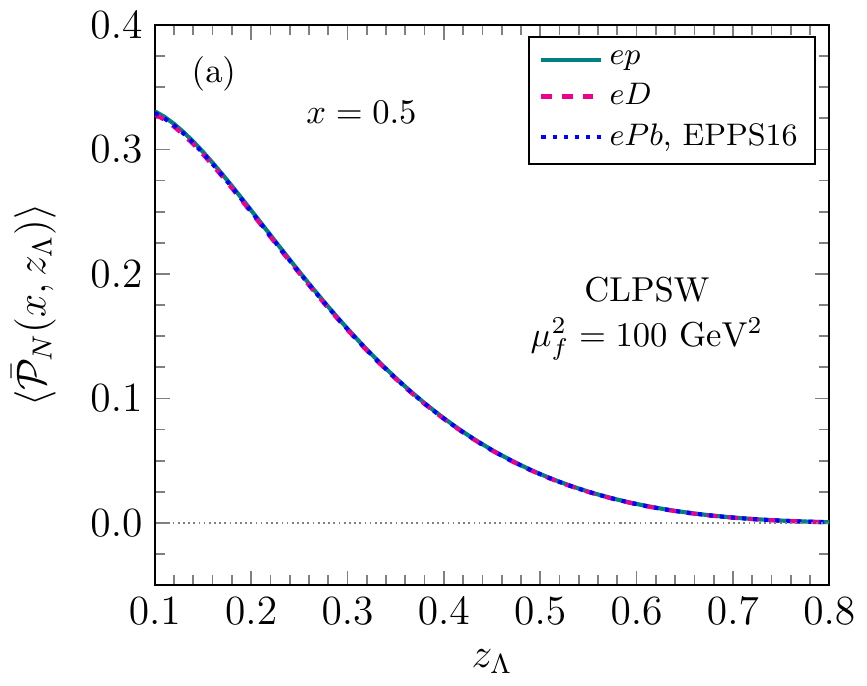}
\includegraphics[width=0.4\textwidth]{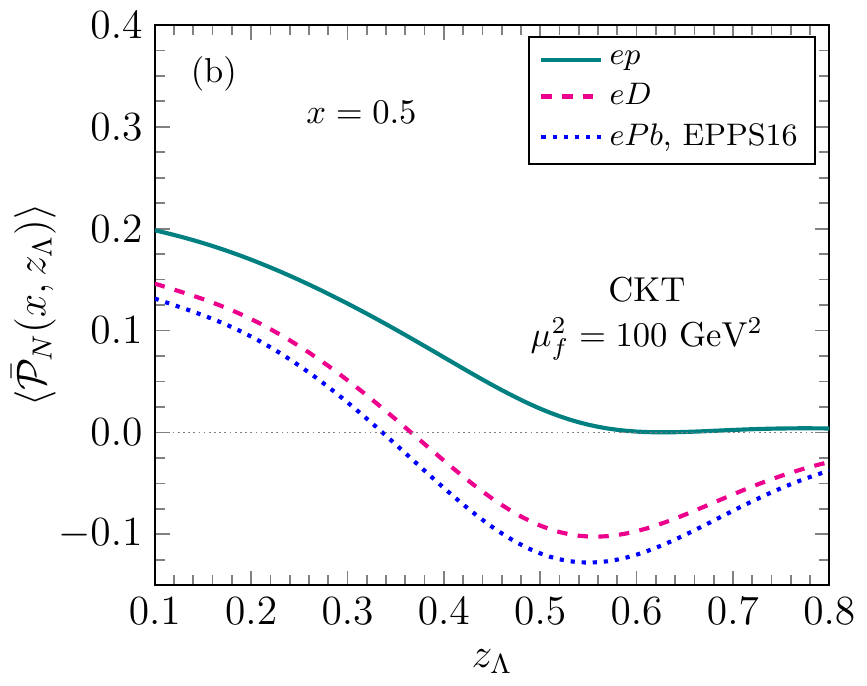}
\includegraphics[width=0.4\textwidth]{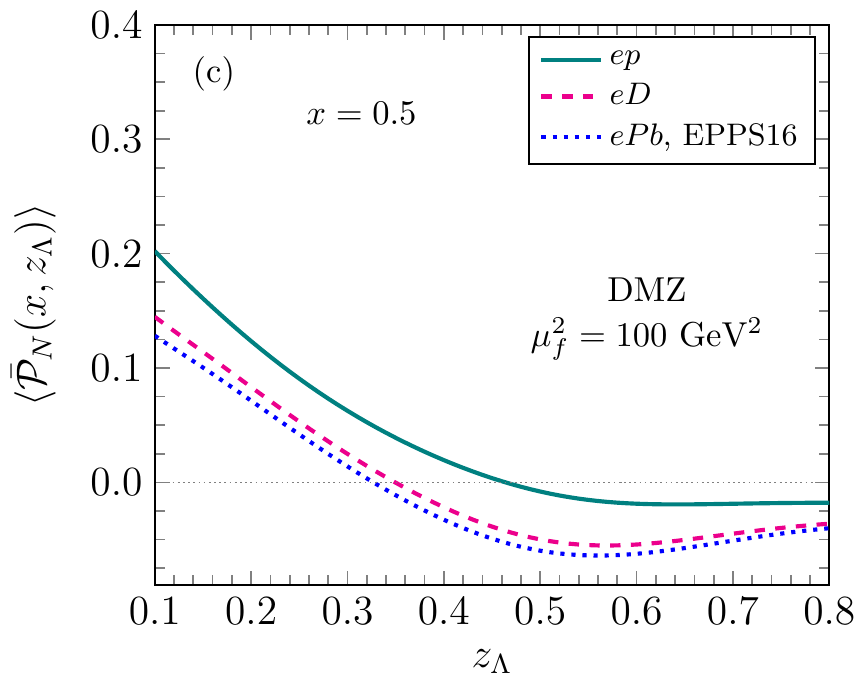}
\caption{Comparison of the transverse polarization of $\Lambda$ in SIDIS with $p$, $D$ and $Pb$ targets obtained using 
different parameterizations~\cite{DAlesio:2020wjq,Callos:2020qtu,Chen:2021hdn} of the polarized FF $D_{1Tq}^{\perp\Lambda}$.  
The unpolarized PDFs, FFs and nuclear PDFs are taken from the CT14~\cite{Dulat:2015mca}, AKK08~\cite{Albino:2008fy}, DSV~\cite{deFlorian:1997zj} 
and EPPS16~\cite{Eskola:2016oht} respectively. }
\label{fig:sidis-tpol-2}
\end{figure}

To this end, we calculate $\langle\bar{\cal P}_N\rangle$ in SIDIS with $p$, $D$ or $Pb$ target  
using three different parameterizations~\cite{Chen:2021hdn,DAlesio:2020wjq,Callos:2020qtu} and show the results obtained in Fig.~\ref{fig:sidis-tpol-2}. 
We see from Fig.~\ref{fig:sidis-tpol-2}, just as expected, at large-$x$, contributions from $u$ and $d$ quarks dominate 
and the isospin symmetric CLPSW parametrization~\cite{Chen:2021hdn} leads to almost identical results for different targets, 
whereas the isospin-symmetry-violating CKT and DMZ parameterizations~\cite{DAlesio:2020wjq,Callos:2020qtu} lead to quite different results for different targets. 
We emphasize in particular that this conclusion is obtained from the $u$ and $d$ quark dominance at large-$x$. 
The uncertainties in FFs may lead to changes in the precise values of the results but the qualitative characteristics remain. 
The measurements of $\langle\bar{\cal P}_N\rangle$ in the large-$x$ region with different nucleus targets provide a clean test of the isospin symmetry in the polarized FF $D_{1Tq}^{\perp\Lambda}$ in general 
and distinguish between different parameterizations~\cite{DAlesio:2020wjq,Callos:2020qtu,Chen:2021hdn} in particular. 

\begin{figure}[htb]
\includegraphics[width=0.4\textwidth]{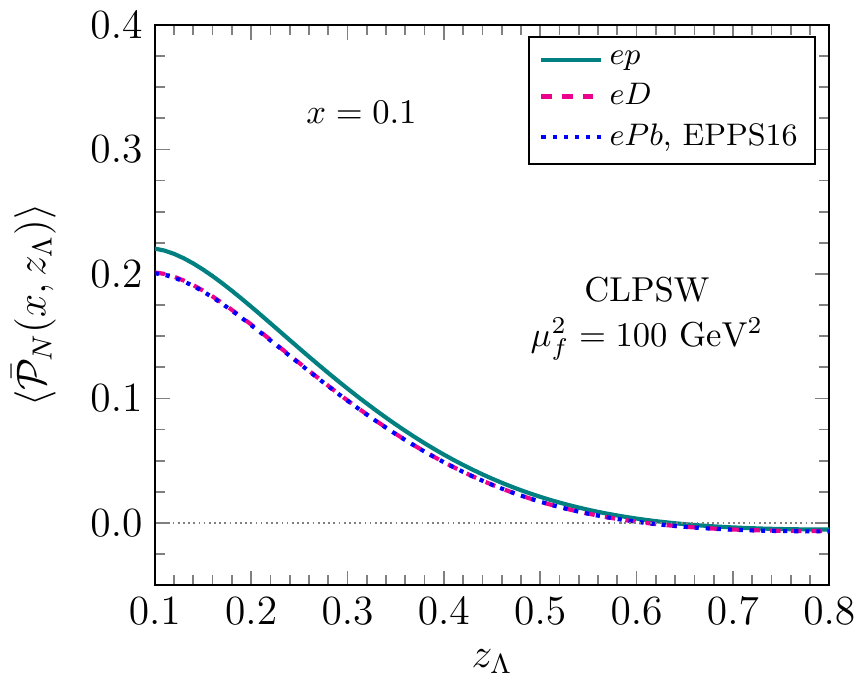}
\caption{The transverse polarization of $\Lambda$ in SIDIS with $p$, $D$ or $Pb$ target at $x=0.1$ with the CLPSW parameterization.}
\label{fig:sidis-tpol-3}
\end{figure}

We also point out that the differences are significant mainly at large $x$. 
In the small-$x$ region, contributions from sea partons dominate. 
The relative weights of different flavors and the flavor dependence of FFs may also lead to a small difference for $\langle\bar{\cal P}_N\rangle$ 
in $ep$ and $eA$ scatterings even in the isospin symmetric CLPSW parameterization as shown in Fig.~\ref{fig:sidis-tpol-3}.

Numerical estimates of other components such as the $\sin2\phi$ weighted longitudinal polarization $\langle \bar{\cal P}_L^{\sin2\phi}\rangle$ 
and the transverse polarizations $\langle \bar{{\cal P}}_T^{\sin2\phi}\rangle$, 
and the $\cos2\phi$ weighted transverse polarization $\langle \bar{{\cal P}}_N^{\cos2\phi}\rangle$ are even more involved and less reliable 
because there is no parameterization or measurement of the corresponding chiral odd spin dependent FF available at all. 
Furthermore, they depend also on the chiral-odd PDF $h_1^\perp (x)$ that is also poorly known. 
Here, we make the following discussions and/or simple rough estimates of the order of magnitude of these three components by 
comparing them with $\langle \bar{\cal P}_N\rangle$. 

(1) From Eqs.~(\ref{eq:PN-SIDIS})-(\ref{eq:PNcos2phi-SIDIS}), we see that 
there is a $y$-dependent factor $B(y)/A(y)={2(1-y)}/[{1+(1-y)^2}]$ for these three components compared to $\langle \bar{\cal P}_N\rangle$. 
This factor is a monotonic function of $y$. It takes unity at $y=0$, decreases with increasing $y$ and takes zero at $y=1$. 

(2) Another calculable factor is $\kappa_i$. They are given by Eqs.~(\ref{eq:kappa1})-(\ref{eq:kappa5}). 
If we take $\Delta_p\sim\Delta_\Lambda\sim m_p\sim M_\Lambda$, we see that these $\kappa_i$'s are all of the same order of magnitude. 

(3) From Eqs.~(\ref{eq:PN-SIDIS})-(\ref{eq:PNcos2phi-SIDIS}), we see also that the major difference between these three components and $\langle \bar{\cal P}_N\rangle$ 
is that the PDF involved is the chiral odd Boer-Mulders function $h_1^\perp (x)$ rather than the number density $f_1(x)$. 

We note that $h_1^\perp (x)$ can be extracted from the Drell-Yan process $pp\to l^+ l^- X$.   
The $\langle \cos2\phi \rangle_{DY}$-asymmetry in this process is roughly proportional to ${h_1^\perp(x_1) h_1^\perp(x_2)}/{f_1(x_1) f_1(x_2)}$. 
Hence, as a very rough estimate of the order of magnitude, we may take roughly $h_{1}^\perp \sim \sqrt{\langle \cos2\phi \rangle_{DY}} f_1$.  
Measurements of $\langle \cos2\phi \rangle_{DY}$ have been carried out by the NuSea collaboration \cite{NuSea:2006gvb}. 
We see that the magnitude of $\langle \cos2\phi \rangle_{DY}$ is about 0.04. 
Therefore, we estimate that the magnitude of $h_1^\perp (x)$ is about 0.2 times that of $f_1(x)$. 
A parameterization of $h_1^\perp (x)$ is available~\cite{Zhang:2008nu,Lu:2009ip} by fitting the NuSea data~\cite{NuSea:2006gvb}, 
and it shows that $h_1^\perp (x)/f_1 (x)$ is indeed of the order of 0.2 in a wide range of $x$. 

(4) For the FFs involved, we note the following: 
Both $H_{1L}^{\perp\Lambda}$ and $D_{1T}^{\perp\Lambda}$ describe induced hyperon polarizations in the fragmentation process. 
While $H_{1L}^{\perp\Lambda}$ describes the longitudinal $\Lambda$ polarization from the fragmentation of a transversely polarized quark, 
$D_{1T}^{\perp\Lambda}$ describes transverse $\Lambda$ polarization from the fragmentation of an unpolarized quark. 
We take $H_{1L}^{\perp\Lambda} (z_\Lambda) \sim H_{1T}^{\perp\Lambda} \sim  D_{1T}^{\perp\Lambda}(z_\Lambda) /z_\Lambda$ as a rough estimate of the order of magnitude,  
where the additional $1/z_\Lambda$ factor comes from the convention taken in the decomposition of the $\hat\Xi$ correlator given by Eq.~(\ref{correlation-matrix-Xi}). 
Similarly, both $H_{1T}^{\Lambda}$ and $G_{1L}^{\Lambda}$ describe spin transfer in the fragmentation process. 
While $H_{1T}^{\Lambda}$ describes the spin transfer in the transversely polarized case, $G_{1L}^{\Lambda}$ describes that in the longitudinally polarized case.  
For a rough estimate of the order of magnitude, we may take $H_{1T}^{\Lambda} \sim G_{1L}^{\Lambda}$. 

\begin{figure}[htb]
\begin{centering}
\includegraphics[width=0.35\textwidth]{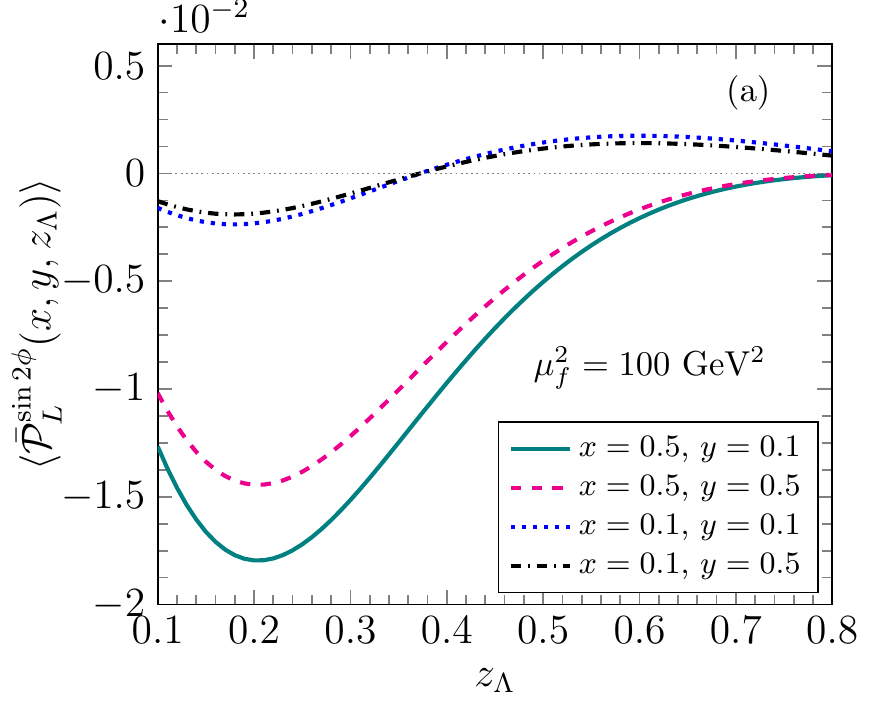}
\includegraphics[width=0.35\textwidth]{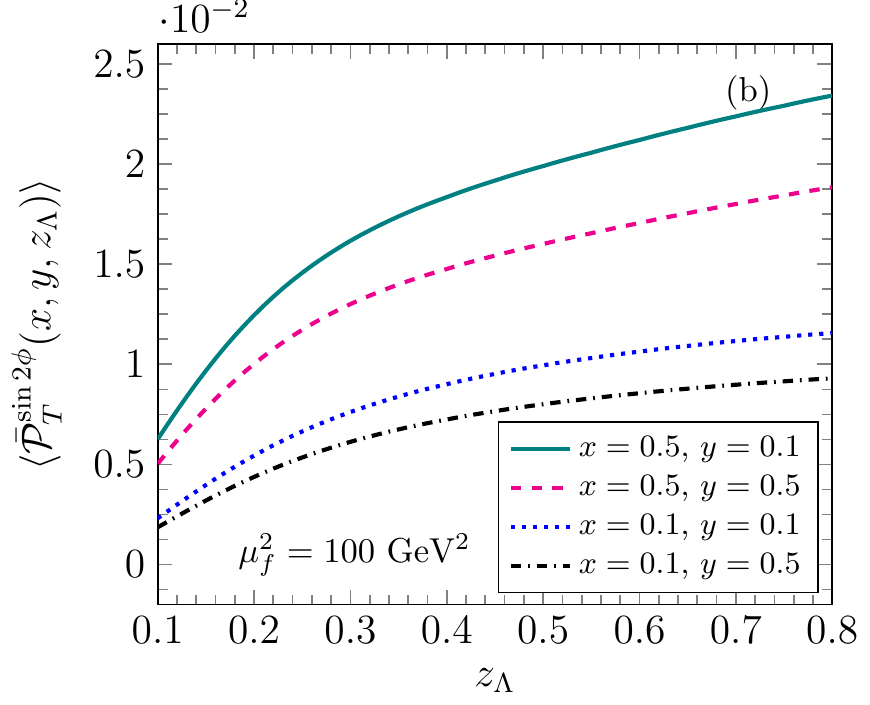}
\includegraphics[width=0.35\textwidth]{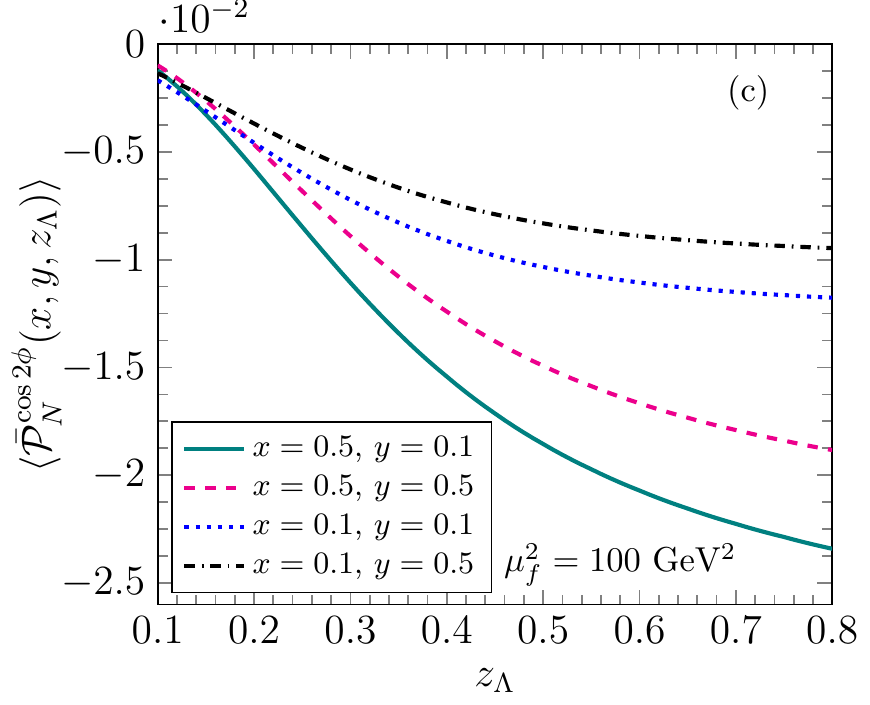}
\caption{Results of rough estimates of (a) $\langle \bar{\cal P}_L^{\sin2\phi}\rangle$, 
(b) $\langle \bar{{\cal P}}_T^{\sin2\phi}\rangle$ and (c) $\langle \bar{{\cal P}}_N^{\cos2\phi}\rangle$ in $e^-p\to e^-\Lambda X$ 
by taking $H_{1L}^{\perp\Lambda} \sim H_{1T}^{\perp\Lambda} \sim D_{1T}^{\perp\Lambda}/z_\Lambda$ and $H_{1T}^{\Lambda} \sim G_{1L}^{\Lambda}$. 
$D_{1T}^{\perp\Lambda}$ is taken from~\cite{Chen:2021hdn}; 
the Boer-Mulders function $h_1^\perp (x)$ is taken from \cite{Lu:2009ip};  
the longitudinal spin transfer $G_{1L}^{\Lambda}$ is taken from~\cite{deFlorian:1997zj}; 
and the unpolarized PDFs and FFs are taken from CT14~\cite{Dulat:2015mca} and DSV~\cite{deFlorian:1997zj} parameterizations. }
\label{fig:PLTN}
\end{centering}
\end{figure}

With points (1) through (4), we expect that the magnitudes of all these three components $\langle \bar{\cal P}_L^{\sin2\phi}\rangle$, 
$\langle \bar{{\cal P}}_T^{\sin2\phi}\rangle$ and $\langle \bar{{\cal P}}_N^{\cos2\phi}\rangle$ should be considerably smaller than that of  $\langle \bar{{\cal P}}_N\rangle$. 
To show the order of magnitude expected more precisely, we present in Fig.~\ref{fig:PLTN} the results of rough estimates  
by taking $H_{1L}^{\perp\Lambda} \sim H_{1T}^{\perp\Lambda} \sim D_{1T}^{\perp\Lambda}/{z_\Lambda}$ and $H_{1T}^{\Lambda} \sim G_{1L}^{\Lambda}$ approximately. 

From Fig.~\ref{fig:PLTN}, we see clearly that these azimuthal-angle-dependent polarizations are indeed expected to be one order of magnitude smaller than $\langle \bar{{\cal P}}_N\rangle$. 
We also see in particular the following qualitative features of results presented in Fig.~\ref{fig:PLTN}. 

(a) The polarizations are in general larger at $x=0.5$ than those at $x=0.1$. 
This is mainly because the Boer-Mulders function given by the parameterization in~\cite{Lu:2009ip} takes the maximum values at intermediate $x$ and becomes smaller at either large or low $x$. 

(b) The polarizations decrease with increasing $y$ as a result of the monotonic behavior of $B(y)/A(y)$.  
In fact the difference between results at different values of $y$ is simply an overall factor given by $B(y)/A(y)$. 

(c) The $z_\Lambda$-dependences of magnitudes of $\langle\bar{\mathcal{P}}_{T}^{\sin 2\phi}\rangle$ and $\langle\bar{\mathcal{P}}_{N}^{\sin 2\phi}\rangle$ are simple 
and increase monotonically with increasing $z_\Lambda$, 
while that of $\langle \bar{\mathcal{P}}_{T}^{\sin 2\phi}\rangle$ is a bit more complicated. 
This feature depends in fact strongly on the approximations that we take for the spin dependent FFs. 
Here, for $\langle\bar{\mathcal{P}}_{T}^{\sin 2\phi}\rangle$ and $\langle\bar{\mathcal{P}}_{N}^{\sin 2\phi}\rangle$,  
we took $H_{1T}^{\Lambda}/D_1^{\Lambda} \sim G_{1L}^{\Lambda}/D_1^{\Lambda} \sim z_\Lambda^\alpha$ 
where $\alpha>0$ in the DSV parameterization~\cite{deFlorian:1997zj}, leading to the monotonic behaviors with increasing $z_\Lambda$.  
For $\langle \bar{\mathcal{P}}_{T}^{\sin 2\phi}\rangle$,  we took $H_{1Lq}^{\perp\Lambda} \sim D_{1Tq}^{\perp\Lambda}/z_\Lambda$ 
and the flavor dependence in the parameterization of $D_{1Tq}^{\perp\Lambda}$ leads to the complicated behavior given by Fig.~\ref{fig:PLTN}a. 

These qualitative features could be used as guides for future measurements.

\section{$\Lambda$ polarizations in $e^+e^-\to\Lambda hX$}

High energy $e^+e^-\to\Lambda hX$ where $\Lambda$ and $h$ are in the opposite hemispheres are quite similar to SIDIS in kinematics. 
Here only TMD FFs are involved. 
In this section, we present the key formulae and results for $\Lambda$ polarizations in $e^+e^-\to\Lambda hX$ 
following the same procedure as that in Sec.~\ref{sec:SIDIS}. 
We consider $e^+e^-$ annihilations in the Belle energy region so that we consider only $e^+e^-\to\gamma^*\to q\bar q\to \Lambda hX$.

\subsection{The kinematic analysis and $\Lambda$ polarizations}

\begin{figure}[htb]
\includegraphics[width=0.45\textwidth]{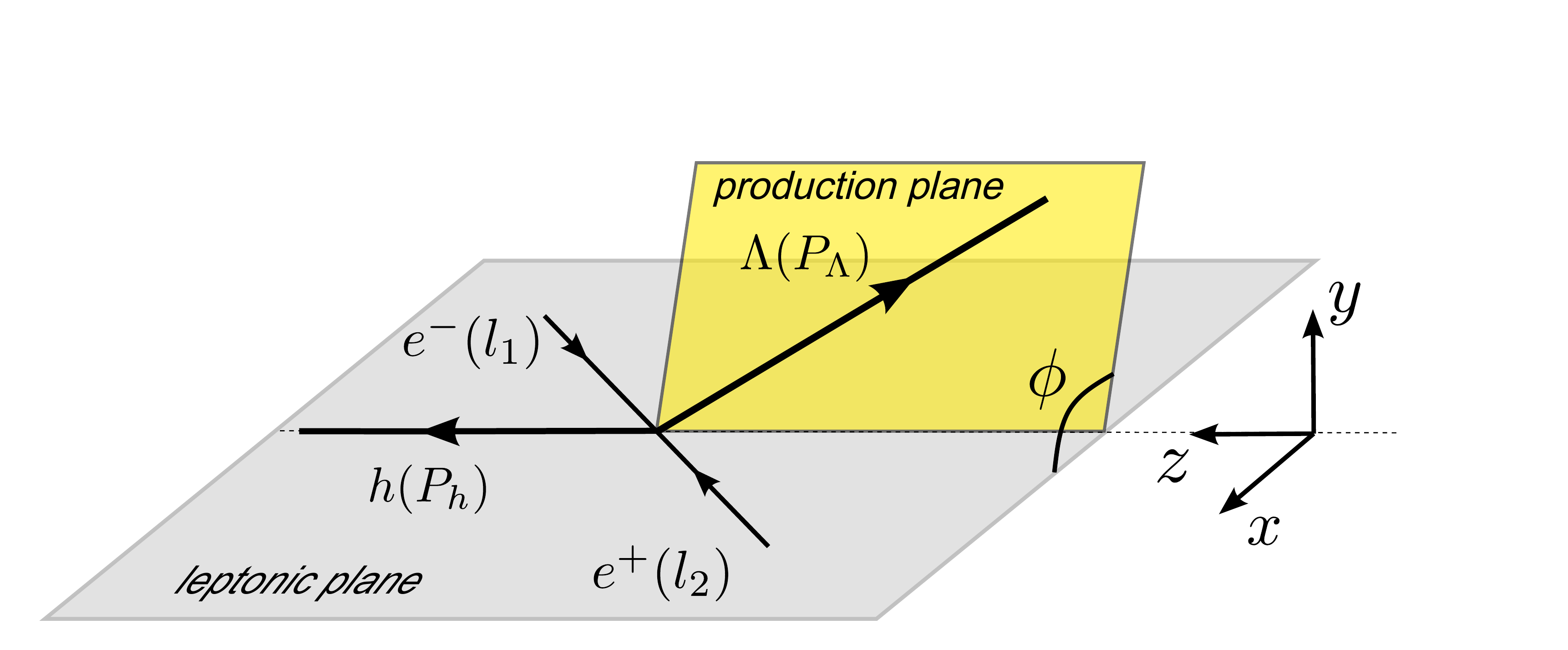}
\caption{Illustrating diagram of the coordinate system for $e^+e^-\to \Lambda h X$. 
The symbol in the bracket denotes the four-momentum of the particle.}
\label{fig:ee-kinematics}
\end{figure}

 We take the center of mass frame of $e^+e^-$ and take the coordinate system as shown in Fig.~\ref{fig:ee-kinematics}.
The differential cross section is given by
\begin{align}
\frac{d\sigma^{e^+e^-}}{dz_h dz_\Lambda dy d^2\bm{P}_{\Lambda\perp}} = \frac{2\pi N_c \alpha_{\rm em}^2}{Q^4} L_{\mu\nu} W^{\mu\nu},
\label{eq:cs-ee-ori}
\end{align}
where $Q^2=q^2=(l_1+l_2)^2$, $z_h = 2P_h \cdot q / Q^2$, $z_\Lambda = 2P_{\Lambda}\cdot q/Q^2$, $y=P_h\cdot l_2 / P_h\cdot q= (1+\cos\theta)/2$ 
with $\theta$ the polar angle between $\bm{l}_1$ and $\bm{P}_h$, $L_{\mu\nu}$ is the leptonic tensor, and $W^{\mu\nu}$ is the hadronic tensor. 

The hadronic tensor for $e^+e^-\to\Lambda hX$ takes exactly the same form as that for the SIDIS $e^-N\to e^-\Lambda X$. 
As a result, the differential cross section take also the same form but with slight differences in the coefficients, i.e., 
\begin{align}
& \frac{d\sigma^{\rm e^+e^-}}{dz_hdz_\Lambda dy d^2\bm{P}_{\Lambda\perp}} = \frac{2\pi N_c\alpha_{\rm em}^2}{Q^2}\Big\{ {\cal A}(y) F_{UU}^T + {\cal B}(y) F_{UU}^L \nonumber \\
& ~~~~~~~~ + {\cal C}(y) \cos\phi F_{UU}^{\cos\phi} + {\cal B}(y) \cos2\phi F_{UU}^{\cos2\phi} \nonumber\\
& ~~~~~~~~ + \lambda_\Lambda \left[ {\cal C}(y) \sin\phi F_{UL}^{\sin\phi} + {\cal B}(y) \sin2\phi F_{UL}^{\sin2\phi} \right] \nonumber\\
& ~~~~~~~~ + S_{\Lambda T} \left[ {\cal C}(y) \sin\phi F_{UT}^{\sin\phi} + {\cal B}(y) \sin2\phi F_{UT}^{\sin2\phi} \right] \nonumber\\
& ~~~~~~~~ + S_{\Lambda N} \Bigl[{\cal A}(y) F_{UT}^T + {\cal B}(y) F_{UT}^L + {\cal C}(y) \cos\phi F_{UT}^{\cos\phi} \nonumber \\
& ~~~~~~~~~~~~~~~~~~  + {\cal B}(y) \cos2\phi F_{UT}^{\cos2\phi}\Bigr] \Big\},
\label{eq:cs-ee-kinematics}
\end{align}
where $ {\cal A}(y) = y^2 + (1-y)^2$, $ {\cal B}(y) = 2y(1-y)$, ${\cal C}(y) =(1-2y)\sqrt{y(1-y)}$, 
and the structure functions $F$'s are scalar functions of $z_h$, $z_\Lambda$, $Q^2$ and $|\bm{P}_{\Lambda\perp}|$. 

From Eq.~(\ref{eq:cs-ee-kinematics}), we obtain immediately the expressions of different components of the $\Lambda$ polarization in terms of structure functions. 
They take the same form as those in Sec.~\ref{sec:SIDISkin} for $e^-N\to e^-\Lambda X$ where the coefficients $A$, $B$ and $C$ are replaced by ${\cal A}$,  ${\cal B}$ and  ${\cal C}$ respectively. 
We do not repeat them here.   

\subsection{Parton model results}

In the parton model, the hadronic tensor for $e^+e^-\to\Lambda hX$ at the LO in pQCD is given by
\begin{align}
W^{\mu\nu} =  & 
e_q^2 \int d^2\bm{p}_T d^2 \bm{p}_{hT} \delta^2 \left(\frac{z_\Lambda}{z_h}\bm{p}_{hT} + \bm{p}_T - \bm{P}_{\Lambda\perp}\right) \nonumber\\
&\times {\rm Tr} \left[ 2\hat\Xi_q^\Lambda (z_\Lambda, \bm{p}_T) \gamma^\mu 2 \hat\Xi_{\bar q}^h (z_h, \bm{p}_{hT}) \gamma^\nu \right], \label{wmunu-ee}
\end{align}
where $\bm{p}_T$ and $\bm{p}_{hT}$ are the transverse momenta of $\Lambda$ and $h$ with respect to the direction of the fragmenting quark,  
a sum over quark flavors and an exchange between $q$ and $\bar q$ are also implicit.
The decomposition of the quark-quark correlator $\hat\Xi_q^\Lambda$ is given by  Eq. (\ref{correlation-matrix-Xi}), 
and that of $\hat\Xi_{\bar q}^h$ is
\begin{align}
  &4\hat\Xi_{\bar q}^h (z_h, \bm{p}_{hT}) = \slashed n_+ D_{1\bar q}^{h} (z_h,p_{hT}) \nonumber\\
  &\phantom{XXX}+ \frac{i [\slashed p_{hT}, \slashed n_+]}{2m_h} H_{1\bar q}^{\perp h} (z_h,p_{hT}). \label{correlation-matrix-Xi-h}
\end{align}
 
By inserting Eqs. (\ref{correlation-matrix-Xi-h}) and (\ref{correlation-matrix-Xi}) into Eq. (\ref{wmunu-ee}), taking traces and contracting with the leptonic tensor, we obtain
\begin{align}
& F_{UU}^T = {\cal I} \left[  D_{1}^h  D_{1}^\Lambda  \right], \\
& F_{UU}^{\cos2\phi} = {\cal I} \left[  \tilde w_2 H_{1}^{\perp h}  H_{1}^{\perp\Lambda}  \right], \\
& F_{UL}^{\sin2\phi} = {\cal I} \left[  \tilde w_2 H_{1}^{\perp h}  H_{1L}^{\perp\Lambda}   \right], \\
& F_{UT}^T = {\cal I} \left[ \tilde{\bar w}_1  D_{1}^h D_{1T}^{\perp\Lambda} \right]/z_\Lambda, \\
& F_{UT}^{\cos2\phi} = {\cal I}[  \tilde w_{3b} H_{1}^{\perp h}    H_{1T}^{\perp\Lambda} ]+{\cal I}[ \tilde w_1 H_{1}^{\perp h} H_{1T}^{\Lambda} ], \\
& F_{UT}^{\sin2\phi} = {\cal I} [ \tilde w_{3a} H_{1}^{\perp h}  H_{1T}^{\perp\Lambda}] - {\cal I}[ \tilde w_1 H_{1}^{\perp h} H_{1T}^{\Lambda}], \\
& F_{UU/UT}^L=F_{UU/UT}^{\cos\phi}=F_{UL/UT}^{\sin\phi}=0,
\end{align}
where ${\cal I}[\tilde wD_1^h D_1^\Lambda] \equiv e_q^2 \int  d^2 \bm{p}_T d^2 \bm{p}_{hT} \delta^2 (\frac{z_\Lambda}{z_h}\bm{p}_{hT}+\bm{p}_T  - \bm{P}_{\Lambda\perp}) \tilde w D_{1\bar q}^h(z_h,p_{hT}) D_{1q}^\Lambda (z_\Lambda,p_T)$ 
and $\tilde w_i$'s are given by 
\begin{align}
& \tilde w_1 = {\hat P_{\Lambda\perp} \cdot p_{hT}}/{m_h}, ~~~~~~ 
 \tilde{\bar w}_1 = -{\hat P_{\Lambda\perp} \cdot p_T}/{M_\Lambda}, \\
& \tilde w_2 =  \Bigl[{2(\hat P_{\Lambda\perp}\cdot p_{hT})(\hat P_{\Lambda\perp}\cdot p_T) + p_{hT}\cdot p_T}\Bigr]/{m_h M_\Lambda}, \\
& \tilde w_{3a} = \Big[2(\hat P_{\Lambda\perp}\cdot p_{hT}) (\hat P_{\Lambda\perp}\cdot p_T)^2 \nonumber\\
&\phantom{XXXX}+(p_{hT}\cdot p_T)(\hat P_{\Lambda\perp}\cdot p_T) \Big]/{m_h M_\Lambda^2}, \\
& \tilde w_{3b} =  \Big[2(\hat P_{\Lambda\perp}\cdot p_{hT})(\hat P_{\Lambda\perp}\cdot p_T)^2 +( p_{hT}\cdot p_T)(\hat P_{\Lambda\perp}\cdot p_T) \nonumber\\
&\phantom{XXXX}  + (\hat P_{\Lambda\perp} \cdot p_{hT})p_T^2 \Big]/{m_h M_\Lambda^2}.
\end{align}

The $\sin n\phi$ and $\cos n\phi$ weighted polarizations are~\cite{ff1} 
\begin{align}
&\langle {\cal P}_N\rangle  = \frac{1}{z_\Lambda}\frac{{\cal I}[\tilde {\bar w}_1 D_1^h D_{1T}^{\perp\Lambda}]}{{\cal I}[D_1^h D_1^\Lambda]}, \label{eq:PNaveepemLT}\\
&\langle {\cal P}_L^{\sin\phi}\rangle =\langle {\cal P}_T^{\sin\phi}\rangle =\langle {\cal P}_N^{\cos\phi}\rangle =0, \label{eq:PsinphiepemLT}\\
&\langle {\cal P}_L^{\sin2\phi}\rangle=\frac{{\cal B}(y)}{2{\cal A}(y)} \frac{{\cal I}[\tilde w_2 H_1^{\perp h}H_{1L}^{\perp\Lambda}]}
        {{\cal I}[D_1^h D_1^\Lambda]},\label{eq:PLsin2phiepemLT}\\
&\langle {\cal P}_T^{\sin2\phi}\rangle=\frac{{\cal B}(y)}{2{\cal A}(y)} \frac{-{\cal I}[\tilde w_1 H_1^{\perp h}H_{1T}^\Lambda ] +{\cal I}[\tilde w_{3a} H_1^{\perp h}H_{1T}^{\perp\Lambda}] }
        {{\cal I}[D_1^h D_1^\Lambda]}, \label{eq:PTsin2phiepemLT}\\
&\langle {\cal P}_N^{\cos2\phi}\rangle=\frac{{\cal B}(y)}{2{\cal A}(y)} \frac{{\cal I}[\tilde w_1 H_1^{\perp h}H_{1T}^\Lambda ]+ {\cal I}[ \tilde w_{3b} H_1^{\perp h}H_{1T}^{\perp\Lambda}]  }
        {{\cal I}[D_1^h D_1^\Lambda]}. \label{eq:PNcos2phiepemLT}
\end{align}

From Eqs.~(\ref{eq:PNaveepemLT})-(\ref{eq:PNcos2phiepemLT}), we see that besides the small differences between the weights $\tilde w_i$'s and $w_i$'s, 
the major differences between different components of $\Lambda$ polarization in $e^+e^-\to\Lambda hX$ and the corresponding components 
given by Eqs.~(\ref{eq:PsinphiSIDISLT})-(\ref{eq:PNcos2phiSIDISLT}) in $e^-N\to e^-\Lambda X$ 
are that PDFs $f_1$ and the Boer-Mulders function $h_1^\perp$ are replaced by the corresponding unpolarized FF $D_1^h$ of $h$ or the Collins function $H_1^{\perp h}$. 
 

\subsection{A rough numerical estimate}

We take the Gaussian ansatz for the transverse momentum dependences, integrate over $|\bm{P}_{\Lambda\perp}|$, and obtain the 
azimuthal angle averaged ${\cal P}_N$ and $\sin n\phi$ and $\cos n\phi$ weighted polarizations as
\begin{align}
&\langle \bar{{\cal P}}_N \rangle 
 =\frac{\sqrt{\pi}}{2}\tilde\kappa_3(\tau)\frac{e_q^2 D_{1\bar q}^h(z_h)D_{1Tq}^{\perp \Lambda}(z_\Lambda)/z_\Lambda}{e_q^2 D_{1\bar q}^h(z_h)D_{1q}^ \Lambda (z_\Lambda)}, \\
&\langle \bar{{\cal P}}_L^{\sin 2\phi} \rangle
      =\frac{\tilde \kappa_1(\tau)}{2} \frac{{\cal B}(y)}{{\cal A}(y)}
      \frac{e_q^2 H_{1\bar q}^{\perp h}(z_h)  H_{1L q}^{\perp\Lambda}(z_\Lambda)} {e_q^2  D_{1\bar q}^h (z_h) D_{1 q}^\Lambda (z_\Lambda) },\\
&\langle \bar{{\cal P}}_T^{\sin 2\phi} \rangle
      =-\frac{\sqrt{\pi}}{4} \tilde \kappa_2(\tau) \frac{{\cal B}(y)}{{\cal A}(y)} \nonumber\\
        &\phantom{X}\times  \frac{e_q^2 H_{1\bar q}^{\perp h} (z_h)\left[-H_{1T q}^\Lambda (z_\Lambda)+H_{1T q}^{\perp\Lambda} (z_\Lambda) \tilde\kappa_4(\tau)\right]}
        {e_q^2 D_{1\bar q}^h (z_h) D_{1 q}^\Lambda (z_\Lambda)},\\
&\langle \bar{{\cal P}}_N^{\cos 2\phi} \rangle
      =-\frac{\sqrt{\pi}}{4}\tilde\kappa_2(\tau) \frac{{\cal B}(y)}{{\cal A}(y)} \nonumber\\
      &\phantom{X}\times \frac{e_q^2 H_{1\bar q}^{\perp h} (z_h)\left[H_{1T q}^\Lambda (z_\Lambda)+H_{1T q}^{\perp\Lambda} (z_\Lambda)\tilde\kappa_5(\tau) \right]}
        {e_q^2 D_{1\bar q}^h (z_h) D_{1 q}^\Lambda (z_\Lambda)},
\end{align}
where $\tau \equiv z_\Lambda/z_h$, $\tilde\kappa_i(\tau)$ is obtained directly from $\kappa_i(z_\Lambda)$ given 
by Eqs.~(\ref{eq:kappa1})-(\ref{eq:kappa5}) by replacing the arguments $z_\Lambda$ by $\tau$, $\Delta_p$ by the Gaussian width $\Delta_h$ 
of TMD FFs for $h$, and $m_p$ by $m_h$ respectively, while $\Delta_\Lambda$ remains. 

We see that we have similar situation as that for SIDIS $e^-N\to e^-\Lambda X$. 
We note that $\langle \bar{{\cal P}}_N \rangle$ has been measured by Belle collaboration~\cite{Belle:2018ttu}
and has been studied phenomenologically in recent papers~\cite{DAlesio:2020wjq,Callos:2020qtu,Chen:2021hdn,Li:2020oto}. 
Here, we make rough estimates of order of magnitude of other three components 
such as $\langle \bar{{\cal P}}_L^{\sin 2\phi} \rangle$, $\langle \bar{{\cal P}}_T^{\sin 2\phi} \rangle$ and $\langle \bar{{\cal P}}_N^{\cos 2\phi} \rangle$ 
by comparing them with $\langle \bar{{\cal P}}_N \rangle$.

Such a rough estimate is quite similar to that we did for $\Lambda$ polarizations in SIDIS $e^-N\to e^-\Lambda X$. 
The kinematic factors are similar and the spin dependent FFs of $\Lambda$ are the same. 
The major difference lies between the Boer-Mulders function $h_1^\perp$ and the Collins function $H_1^{\perp h}$. 
While $h_1^\perp$ can be extracted from Drell-Yan process, $H_1^{\perp h}$ can be extracted similarly 
from $\langle \cos2\phi\rangle_{hh}$ in $e^+e^-\to h_1h_2+X$, which is roughly $\langle \cos2\phi\rangle_{hh}\sim H_1^{\perp h_1}(z_1)H_1^{\perp h_2}(z_2)/D_1^{h_1}(z_1)D_1^{h_2}(z_2)$. 
This $\langle\cos2\phi\rangle_{hh}$ has been measured in the Belle~\cite{Belle:2005dmx,Belle:2008fdv}, 
BaBar~\cite{BaBar:2013jdt} and BESIII~\cite{BESIII:2015fyw} experiments. 
A parameterization of $H_1^{\perp h}$ for pions is also available~\cite{Anselmino:2015sxa}. 

\begin{figure}[htb]
\begin{centering}
\includegraphics[width=0.35\textwidth]{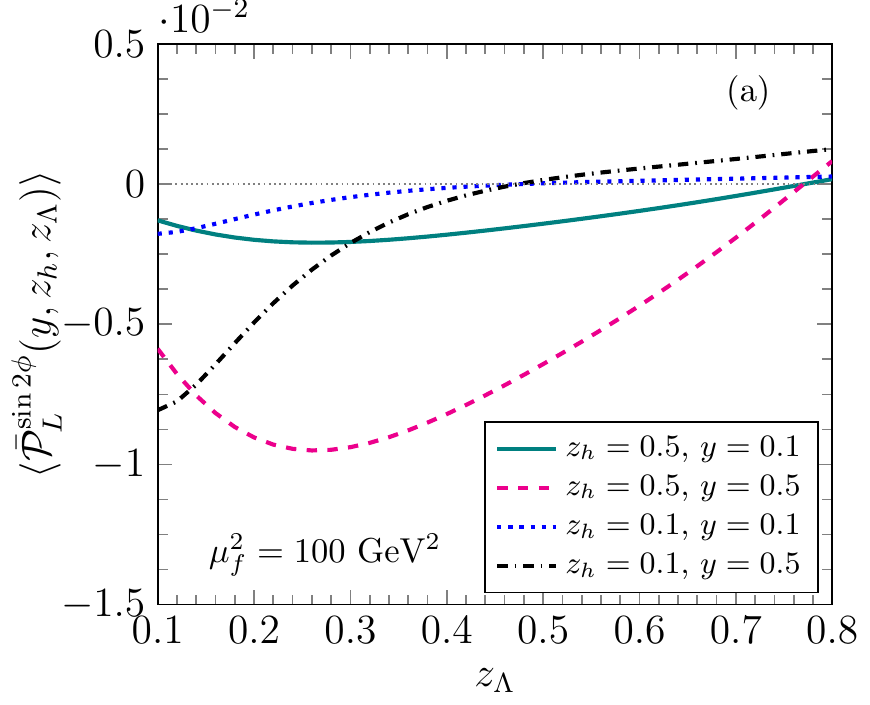}
\includegraphics[width=0.35\textwidth]{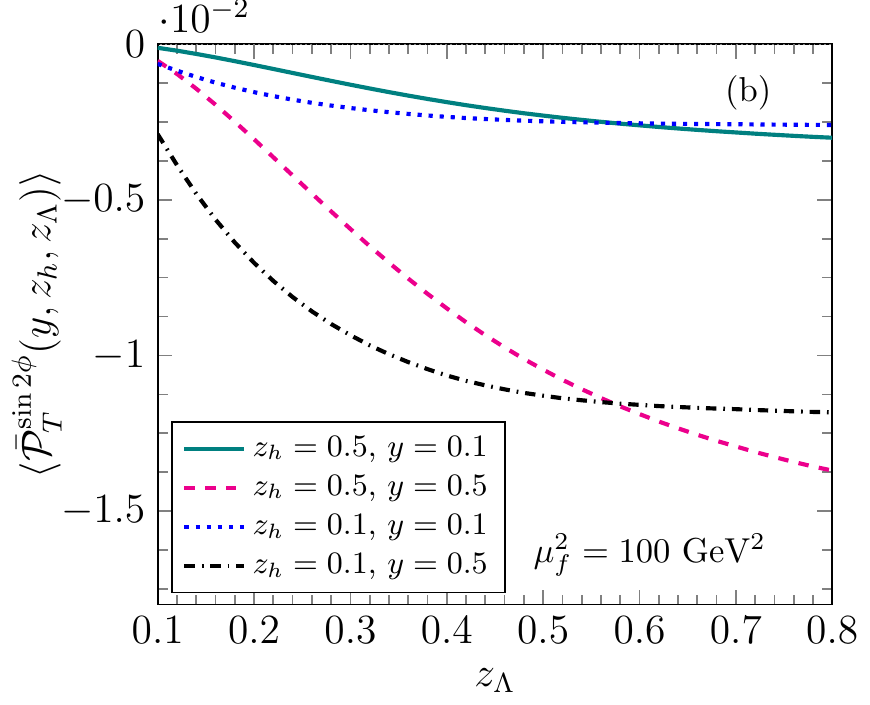}
\includegraphics[width=0.35\textwidth]{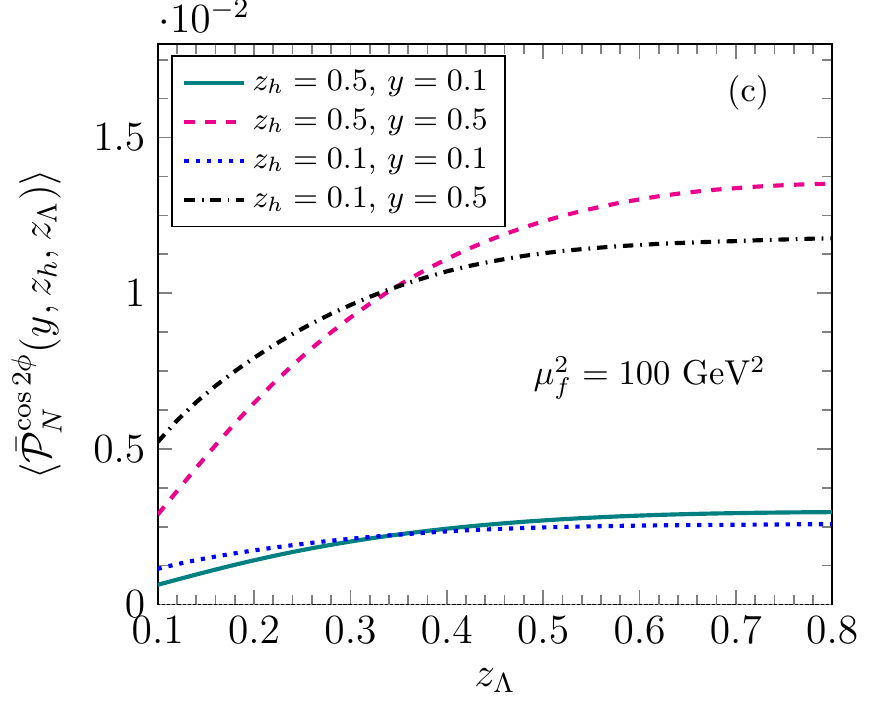}
\caption{Results of rough estimates of (a) $\langle \bar{\cal P}_L^{\sin2\phi}\rangle$, (b) $\langle \bar{{\cal P}}_T^{\sin2\phi}\rangle$ 
and (c) $\langle \bar{{\cal P}}_N^{\cos2\phi}\rangle$ in $e^+e^-\to\Lambda \pi^+ X$ 
by taking $H_{1L}^{\perp\Lambda} \sim H_{1T}^{\perp\Lambda} \sim D_{1T}^{\perp\Lambda}/z_\Lambda$ and $H_{1T}^{\Lambda} \sim G_{1L}^{\Lambda}$. 
The Collins function for $\pi^+$ is taken from \cite{Anselmino:2015sxa}; $D_{1T}^{\perp\Lambda}$ is taken from~\cite{Chen:2021hdn}; 
the longitudinal spin transfer $G_{1L}^{\Lambda}$ is taken from~\cite{deFlorian:1997zj}; 
and the unpolarized FFs are taken from the DSV~\cite{deFlorian:1997zj} and DEHSS~\cite{deFlorian:2014xna} parameterizations. }
\label{fig:PLTNee}
\end{centering}
\end{figure}

With the same assumptions and/or approximations for the polarized FFs of $\Lambda$ such as 
$H_{1L}^{\perp\Lambda} \sim H_{1T}^{\perp\Lambda} \sim D_{1T}^{\perp\Lambda}/z_\Lambda$ and $H_{1T}^{\Lambda} \sim G_{1L}^{\Lambda}$, 
we make similar rough estimates for $\langle \bar{\cal P}_L^{\sin2\phi}\rangle$, 
$\langle \bar{{\cal P}}_T^{\sin2\phi}\rangle$ and $\langle \bar{{\cal P}}_N^{\cos2\phi}\rangle$ in $e^+e^-\to \Lambda\pi^+ X$. 
The results obtained are presented in Fig.~\ref{fig:PLTNee}. 

From Fig.~\ref{fig:PLTNee}, we see that the magnitudes of these polarizations are about $1\%$ and should be measurable in experiments such as Belle II. 
We also see similar qualitative features such as those discussed above in connection with Fig.~\ref{fig:PLTN} for $e^-N\to e^-\Lambda X$. 
Such features might be useful to guide future measurements.

\section{Summary and outlook}

In this paper, we perform a phenomenological study of different components of $\Lambda$ hyperon polarizations in lepton induced reactions such as $e^-N\to e^-\Lambda X$ and $e^+e^-\to\Lambda hX$. 
We started with presenting the general form of cross sections in terms of structure functions obtained from a general kinematic analysis.  
The results show clearly that the produced $\Lambda$ hyperons are in general polarized along the longitudinal direction, the normal direction of the production plane, and the transverse direction in the production plane. 
However all these three components of polarizations depend on the azimuthal angle $\phi$ and two of them vanish if averaged over the whole phase space of $\phi$ and cannot be measured in the conventional way. 
We thus propose either to measure them in different quadrants separately or to measure the $\sin n\phi$ or $\cos n\phi$ weighted polarizations such as 
$\langle{\cal P}_L^{\sin2\phi}\rangle$, $\langle{\cal P}_T^{\sin2\phi}\rangle$ and $\langle{\cal P}_N^{\cos2\phi}\rangle$. 

At the leading twist and LO in pQCD, these different components are all convolutions of TMD PDFs or FFs with polarized FFs. 
While $\langle{\cal P}_N\rangle$ is a convolution of $f_1$ or $D_1$ with $D_{1T}^{\perp\Lambda}$, 
the other three components $\langle{\cal P}_L^{\sin2\phi}\rangle$, $\langle{\cal P}_T^{\sin2\phi}\rangle$ and $\langle{\cal P}_N^{\cos2\phi}\rangle$ 
are all convolutions of chiral odd PDF $h_1^\perp$ or FF $H_1^{\perp h}$ with chiral odd FFs such as $H_{1L}^\perp$, $H_{1T}$ and $H_{1T}^\perp$. 

Using parameterizations available, we made a rough estimate of magnitudes of these $\Lambda$ polarizations based on some radical approximations.  
The numerical results obtained provide a guide for future experiments. 
We show in particular that measuring $\langle{\cal P}_N\rangle$ in SIDIS with different nuclear targets in future EIC experiments provide a good test of the isospin symmetry of FFs. 
Though much smaller than $\langle{\cal P}_N\rangle$, 
the other three components $\langle{\cal P}_L^{\sin2\phi}\rangle$, $\langle{\cal P}_T^{\sin2\phi}\rangle$ and $\langle{\cal P}_N^{\cos2\phi}\rangle$ can be measured 
at Belle II and future EIC with high luminosities. 
Such measurements provide an effective way to study the chiral-odd FFs such as $H_{1L}^\perp$, $H_{1T}$ and $H_{1T}^\perp$ in the unpolarized SIDIS and $e^+e^-$-annihilation.

\begin{acknowledgments}

This work was supported in part by the National Natural Science Foundation of China (approval Nos. 11890713, 12005122, 11890710, 11947055, 11505080) and Shandong Province Natural Science Foundation Grant No. ZR2018JL006 and ZR2020QA082.

\end{acknowledgments}

\appendix

\section{Measurements of longitudinal and transverse polarizations of hyperons}
\label{sec:app:measurement}
In this appendix, we discuss the method to measure $\sin n\phi$ and $\cos n\phi$ weighted polarizations, $\langle {\cal P}_{j}^{\sin n\phi}\rangle$ and $\langle {\cal P}_{j}^{\cos n\phi}\rangle$ where $j$ specifies the direction of the polarization. 
To be explicit, we take SIDIS $e^-N\to e^-\Lambda  X$ as an example and it can be easily extended to $e^+e^-$ annihilation.

The $\Lambda$ polarization along any specific direction can be measured via the angular distribution of its decay products in the parity violating weak decay, e.g., $\Lambda\to p+\pi^-$. The normalized $\cos\theta_j^*$-distribution in the $\Lambda$-rest frame is
\begin{align}
\frac{dN}{d\cos\theta_j^*} = \frac{1}{2}\left(1 + \alpha {\cal P}_{j} \cos\theta_j^*\right),
\label{decay}
\end{align}
where $\theta_j^*$ is the angle between the proton momentum in the rest frame of $\Lambda$ and the specific direction $j$, $\alpha$ is the decay parameter of $\Lambda$, and ${\cal P}_{j}$ is the $\Lambda$ polarization along this specific direction. 

We consider the normalized two-variable joint distribution function, $f(\phi, \cos\theta_j^*)$, of final state protons from decays of $\Lambda$ hyperons produced 
in $e^-N\to e^-\Lambda  X$ at the azimuthal angle $\phi$. 
Such a joint distribution is given by the product of the cross section to produce $\Lambda$ hyperons multiplied by the distribution given by Eq.~(\ref{decay}), i.e., 
\begin{align}
f(\phi,\cos\theta_j^*) = & \frac{1}{\sigma_{\rm tot}} \int dxdydz_\Lambda |\bm{P}_{\Lambda\perp}|d|\bm{P}_{\Lambda\perp}| \frac{d\sigma_{UU}^{\rm SIDIS}}{dxdydz_\Lambda d^2\bm{P}_{\Lambda\perp}} \nonumber\\
& \times \frac{dN}{d\cos\theta_j^*},
\end{align}
where $\sigma_{\rm tot}$ is the total cross section in the kinematical region specified in experiments. 
The normalization of this distribution function follows $\int d\phi d\cos\theta_j^* f(\phi,\cos\theta_j^*) =1$. 
It is then straightforward to find 
\begin{align}
&
\langle {\cal P}_{j}^{\sin n\phi} \rangle = \frac{3}{\alpha} \langle \sin n\phi \cos\theta_j^* \rangle
\nonumber\\
& \phantom{XXX} \equiv \frac{3}{\alpha} \int d\phi d\cos\theta_j^* ~ f(\phi,\cos\theta_j^*) \sin n\phi \cos\theta_j^*,
\\
&\langle {\cal P}_{j}^{\cos n\phi} \rangle =  \frac{3}{\alpha} \langle \cos n\phi \cos\theta_j^* \rangle
\nonumber\\
& \phantom{XXX} \equiv \frac{3}{\alpha} \int d\phi d\cos\theta_j^* ~ f(\phi,\cos\theta_j^*) \cos n\phi \cos\theta_j^*.
\end{align}

\end{document}